\documentclass[a4paper,fleqn]{cas-dc}
\ifdefined\pdfminorversion\pdfminorversion=7\relax\fi
\usepackage{pifont}
\usepackage{algorithm}
\usepackage{algpseudocode}
\usepackage{float}
\usepackage{flushend}
\usepackage[numbers]{natbib}
\usepackage{placeins}
\usepackage{algpseudocode}
\usepackage{booktabs}
\usepackage{tabularx}
\usepackage{array}

\def\tsc#1{\csdef{#1}{\textsc{\lowercase{#1}}\xspace}}
\tsc{WGM}
\tsc{QE}
\tsc{EP}
\tsc{PMS}
\tsc{BEC}
\tsc{DE}

\begin{document}
\let\WriteBookmarks\relax
\def\floatpagepagefraction{1}
\def\textpagefraction{.001}
\shorttitle{GRASP}
\shortauthors{Rui Xu et~al.}
\renewcommand{\printorcid}{}
\ExplSyntaxOn
\cs_set:Npn \__first_footerline:
  { { \small \sffamily \__short_authors: :~ \rmfamily \itshape Preprint } }
\ExplSyntaxOff

\title [mode = title]{GRASP: Query-Adaptive Graph Retrieval framework with Cross-Attention Scoring and Path Sufficiency Verification for Question Answering}                      



\author[1,2]{Rui Xu}
\author[2]{Le Wang}
\cormark[1]
\author[3]{Tianyue Wang}

\affiliation[1]{
    organization={Department of Biostatistics, Karolinska Institutet},
    city={Stockholm},
    country={Sweden}
}

\affiliation[2]{
    organization={School of Electrical Engineering and Computer Science, KTH Royal Institute of Technology},
    city={Stockholm},
    country={Sweden}
}

\affiliation[3]{
    organization={Department of Industrial and Operations Engineering, University of Michigan},
    city={Ann Arbor},
    country={USA}
}

\cortext[cor1]{Corresponding author}

\begin{abstract}
Retrieval quality is a key factor in the reliability of retrieval-augmented generation (RAG), particularly when retrieval is performed over structured knowledge graphs. Graph-based retrieval requires decisions about candidate-edge scoring, traversal termination, and the sufficiency of the retrieved evidence. We propose GRASP, a query-adaptive graph retrieval framework that jointly addresses these decisions during the retrieval process. A cross-attention scorer models fine-grained interactions between individual query tokens and candidate edges, providing query-specific signals for edge selection. A learned stopping gate conditions traversal termination on the query, traversal state, and hop depth, allowing the retrieval depth to adapt to the current retrieval state. A sufficiency classifier further evaluates the assembled path before generation and triggers re-retrieval when the evidence is predicted to be incomplete. We evaluate GRASP on three biomedical question-answering benchmarks with different graph scales and answer lengths, using biomedical question answering as a representative setting for evaluating graph-based RAG retrieval. GRASP consistently improves over the strongest learned baseline across the evaluated retrieval and generation metrics. The unsupported-claim rate is reduced by 8.9 to 13.2 percentage points across the three benchmarks, with the largest reduction on the benchmark with the longest generated answers. The retrieval gains are more pronounced under higher candidate density, while ablation results indicate that the three components provide complementary improvements.


\end{abstract}





\maketitle

\section{Introduction}
\label{sec:intro}

Retrieval-augmented generation (RAG) grounds large language model (LLM)
outputs in externally retrieved evidence, making retrieval quality an
important factor in the reliability of knowledge-intensive tasks
\citep{Lewis2020,Shuster2021}. Recent work has explored query
reformulation, retrieval enhancement, and adaptive retrieval to improve
the evidence supplied to the generator \citep{Khot2021,Trivedi2023}.
When retrieval is performed over a knowledge graph (KG), however,
retrieval quality involves more than identifying individually relevant
facts. The retriever must construct a connected path that provides
sufficient evidence for answering the query. In particular, the
relevance of retrieved edges does not necessarily imply that the
assembled path covers the evidence required by the answer.

Graph-based retrieval represents knowledge through entities and typed
relations, enabling multi-hop retrieval when a query requires several
connected facts \citep{Bodenreider2004,Himmelstein2017,JiKGSurvey2022}.
This setting involves three related but distinct decisions: selecting
relevant candidate edges, determining when to terminate traversal, and
assessing whether the resulting path provides sufficient evidence for
answering the query. These decisions address different aspects of
retrieval. Candidate-edge scoring estimates the relevance of individual
retrieval actions, while traversal termination determines the depth of
the current search. Neither decision, on its own, establishes whether
the accumulated path provides adequate support for the answer. A
retrieval process may therefore terminate with a path containing
relevant edges but lacking one or more necessary relations. If this
incomplete path is passed directly to the generator, the resulting
answer may contain claims that are not supported by the retrieved
evidence.

Existing approaches address different aspects of graph retrieval.
Embedding-based scoring and graph-aware representations support
candidate selection, while learned traversal policies condition path
construction on the query and traversal history
\citep{Xiong2017,Das2018,Vaswani2017}. Other methods investigate
graph-based evidence organization, subgraph or path selection, and
iterative retrieval
\citep{Edge2024,Guo2024,Jimenez2024,Zhu2025,Chen2025,He2024,Hu2024}.
These approaches provide complementary strategies for selecting and
organizing graph evidence. However, candidate relevance and traversal
termination do not necessarily establish path-level evidence
sufficiency. In particular, when a retrieved path is incomplete, simply
terminating traversal or passing the path to the generator does not
resolve the missing evidence. This motivates a retrieval mechanism that
not only constructs a path but also evaluates its sufficiency before
generation and attempts to correct insufficient retrieval results.

To address this problem, we propose GRASP, a query-adaptive graph retrieval framework comprising five stages, labelled A to E. Stage A encodes the input query into representations that guide the subsequent graph retrieval. Stage B constructs a candidate evidence path through query-conditioned edge scoring and adaptive traversal termination. Stage C evaluates whether the assembled path is structurally sufficient to support answering the query. Based on this assessment, Stage D determines whether the retrieved path should be accepted or another retrieval attempt should be initiated. When the path is predicted to be insufficient, GRASP initiates a bounded re-retrieval procedure: it identifies a low-confidence retrieval action using its hop-level score margin, excludes that action in subsequent retrieval attempts, and restarts traversal from the entry entity to seek an alternative path. Once a path is accepted, Stage E uses the retrieved evidence as input to the language model to generate the clinical answer. Thus, Stages A and E provide the query-encoding and answer-generation interfaces of GRASP, respectively, while the central methodological contribution lies in coordinating path-level sufficiency assessment with targeted retrieval correction before generation, rather than relying solely on edge relevance and traversal termination.

Stage B employs a cross-attention scorer to model interactions between
individual query tokens and candidate edges, together with a learned
stopping gate conditioned on the query, traversal state, and hop depth.
The scorer provides fine-grained signals for selecting candidate
actions, while the stopping gate controls the depth of the current
traversal. Stage C addresses a separate question: whether the path
assembled by these retrieval decisions provides sufficient structural
support for the query. It encodes the ordered retrieval path and
combines path-level information with retrieval confidence statistics,
including edge scores, hop-level score margins, candidate uncertainty,
and path length. The resulting sufficiency estimate is used to decide
whether the path can proceed to generation or whether another retrieval
attempt is needed. Stage B is trained using hop-level supervision,
whereas Stage C is trained separately using confidence statistics
produced by the converged retrieval model. At inference time, the two
stages interact through a retrieval verification and correction loop.
Importantly, this mechanism treats insufficient retrieval as a
retrieval-level problem to be addressed before generation, rather than
relying on the generator to compensate for missing path evidence.

We evaluate GRASP on three biomedical question-answering benchmarks:
RiTeK, MedKGent-KG, and a PMC-based biomedical corpus
\citep{RiTeK2025,MedKGent2025,PMC2024}. The benchmarks cover different
graph scales, retrieval settings, and answer-generation requirements.
Across the evaluated benchmarks, GRASP improves over the strongest
learned baseline on the reported retrieval and generation metrics. The
unsupported claim rate decreases by 8.9 to 13.2 percentage points, with the largest
reduction observed on the benchmark with the longest generated answers.
Retrieval gains are also more pronounced in higher-density candidate
settings, although this observation alone does not establish a causal
relationship between graph density and performance. Ablation results
indicate that the three components contribute complementary
improvements.

The main contributions of this paper are summarized as follows:
\begin{enumerate}
    \item A query-adaptive graph retrieval framework, GRASP, is proposed
    to coordinate candidate-edge selection, traversal termination, and
    path-sufficiency verification. The framework distinguishes the
    relevance of individual retrieval actions from the sufficiency of
    the assembled evidence path, addressing the limitation that a
    traversal may terminate with relevant but incomplete evidence.
    Experiments on three biomedical question-answering benchmarks
    demonstrate consistent improvements over the strongest learned
    baseline across the evaluated retrieval and generation metrics.

    \item A cross-attention-based edge scoring mechanism coupled with
    a learned stopping gate is developed to support fine-grained
    query-edge matching and adaptive traversal depth. The scorer
    explicitly models token-level query-edge interactions, while the
    stopping gate conditions termination on the query and current
    traversal state. Ablation results demonstrate the complementary
    contributions of edge scoring and adaptive stopping to retrieval
    performance.

    \item A path-sufficiency verification and retrieval correction
    mechanism is introduced to assess assembled evidence before answer
    generation. Unlike a stopping decision based only on traversal
    state, this mechanism evaluates the retrieved path as a whole and
    identifies whether its structural evidence is predicted to be
    insufficient. When insufficiency is detected, GRASP uses
    hop-level confidence margins to exclude a low-confidence retrieval
    action and initiates a bounded alternative retrieval attempt.
    Experimental results show that GRASP reduces the unsupported-claim
    rate by 8.9 to 13.2 percentage points across the evaluated benchmarks, supporting
    the benefit of incorporating explicit path-sufficiency assessment
    and retrieval correction into graph-based retrieval.
\end{enumerate}

The remainder of this paper is organized as follows. Section~\ref{sec:related}
reviews related work on RAG architectures, graph-based retrieval, and
learned traversal. Section~\ref{sec:problem} formulates the retrieval
problem and introduces the training objectives. Section~\ref{sec:method}
describes the GRASP framework and its inference procedure.
Section~\ref{sec:experiments} presents the experimental results,
including the ablation studies. The final section discusses the
findings and concludes the paper.

\begin{table*}[t]
\centering
\small
\caption{Abbreviations, datasets, models, and technical terms used in this paper.}
\label{tab:abbreviations}

\renewcommand{\arraystretch}{1.18}
\setlength{\tabcolsep}{6pt}

\begin{tabularx}{\textwidth}{
    @{}
    >{\raggedright\arraybackslash}p{0.21\textwidth}
    >{\raggedright\arraybackslash}X
    >{\raggedright\arraybackslash}p{0.17\textwidth}
    @{}
}
\toprule
\textbf{Abbreviation / Name}
&
\textbf{Full Name / Description}
&
\textbf{Category}
\\
\midrule

RAG
& Retrieval-Augmented Generation \citep{Lewis2020}
& General term \\

LLM
& Large Language Model \citep{Brown2020}
& General term \\

KG
& Knowledge Graph \citep{Bodenreider2004,Himmelstein2017}
& General term \\

NLI
& Natural Language Inference, used for claim-level evidence verification \citep{Min2023,Ji2023}
& Evaluation method \\

UMLS
& Unified Medical Language System, a biomedical terminology and knowledge resource \citep{Bodenreider2004}
& Biomedical resource \\

\midrule

PMC
& PubMed Central, a full-text archive of biomedical and life sciences literature \citep{PMC2024}
& Dataset / Corpus \\

RiTeK
& A dataset for complex reasoning over textual knowledge graphs in medicine \citep{RiTeK2025}
& Dataset \\

MedKGent-KG
& A large-scale biomedical knowledge graph for question answering, released by ShowerMaker \citep{ShowerMaker2024}
& Dataset \\

\midrule

MINERVA
& A multi-hop knowledge graph reasoning model based on reinforcement learning \citep{Das2018}
& Baseline model \\

MINERVA-PubMedBERT
& MINERVA equipped with a PubMedBERT encoder \citep{Das2018,Gu2021}
& Baseline model \\

PubMedBERT
& A biomedical language model pretrained on PubMed abstracts and full-text articles \citep{Gu2021}
& Language model \\

G-Retriever
& A graph-based retrieval-augmented generation method for textual graph understanding and question answering \citep{He2024}
& Baseline model \\

GNN-RAG
& A graph neural network-based retrieval method for multi-hop reasoning with large language models \citep{mavromatis2025gnnrag}
& Baseline model \\


\bottomrule
\end{tabularx}
\end{table*}

\section{Related Work}
\label{sec:related}

This section reviews three research directions relevant to graph-based RAG: RAG architectures and evidence verification, graph-based retrieval for structured knowledge, and learned traversal for multi-hop retrieval.

\subsection{RAG Architectures and Evidence Verification}
\label{sec:related-rag}

RAG conditions generation on external evidence rather than relying solely on parametric knowledge \citep{Lewis2020,Shuster2021}. Existing architectures differ in how retrieval is trained and how retrieved information is incorporated into generation. REALM jointly trains retrieval with language model pretraining \citep{Guu2020}, whereas Fusion-in-Decoder encodes passages independently and combines their information during decoding \citep{Izacard2021}. These approaches represent different strategies for integrating retrieved evidence into language generation.

A related line of research evaluates the factual support of generated responses. Claim-level evaluation methods such as FActScore decompose generated text into atomic facts and assess their support against external sources \citep{Min2023}. This line of work focuses on the grounding of generated claims, distinguishing it from the retrieval time question of whether the evidence collected for a query is sufficient before generation. The two concerns are related but operate at different stages of the RAG pipeline.

\subsection{Graph-Based Retrieval for Structured Knowledge}
\label{sec:related-graph}

Graph-based retrieval represents knowledge through entities and typed relations, allowing retrieval methods to exploit connections among facts. Biomedical resources such as UMLS and curated biomedical networks provide structured knowledge for retrieval and reasoning \citep{Bodenreider2004,Himmelstein2017}. Graph neural networks, including graph convolutional and attention-based models, provide mechanisms for propagating information over graph structures and learning representations of related entities \citep{Kipf2017,Velickovic2018}.

Recent graph-based RAG methods differ in how they organize graph information and select evidence for generation. GraphRAG, LightRAG, and HippoRAG use graph-based representations to organize related information for retrieval and generation \citep{Edge2024,Guo2024,Jimenez2024}, while KG2RAG incorporates knowledge-graph structure into evidence organization \citep{Zhu2025}. Other methods focus more directly on retrieving graph structures or paths for language model reasoning. G-Retriever retrieves graph structures as context, whereas PathRAG selects path-based evidence \citep{He2024,Chen2025}. GRAG and QMKGF investigate graph retrieval under different graph structures and retrieval settings \citep{Hu2024,Wei2026}. Graph foundation models and language model-based approaches have also been explored for graph retrieval, including GFM-RAG and ByoKG-RAG \citep{Luo2025,Mavromatis2025ByoKG}.

Taken together, these methods cover several aspects of graph-based evidence retrieval, including graph organization, representation learning, subgraph selection, and path-based retrieval. Their retrieval strategies vary with the graph representation and task setting, providing different ways to construct structured context for downstream generation.

\subsection{Learned Traversal for Multi-Hop Retrieval}
\label{sec:related-traversal}

Learned traversal methods formulate graph retrieval as a sequence of decisions conditioned on the query and the current traversal state. MINERVA models multi-hop reasoning as a Markov decision process and trains a recurrent policy with REINFORCE to select relations and entities \citep{Das2018,Sutton2018}. DeepPath learns relation paths through policy gradients \citep{Xiong2017}, while M-Walk incorporates Monte Carlo Tree Search to improve exploration during path construction \citep{Shen2018}. GNN-RAG combines learned traversal with graph neural network reasoning over retrieved subgraphs \citep{mavromatis2025gnnrag}. These methods illustrate different approaches to query dependent path construction and graph reasoning.

Two related aspects of traversal are candidate selection and termination. Candidate selection methods must distinguish edges that may be semantically similar but differ in their relevance to the current query. Attention mechanisms model interactions between query tokens and candidate information \citep{Vaswani2017}, while retrieval and reranking studies have explored query evidence alignment to improve evidence selection \citep{Khot2021,Trivedi2023}. For traversal termination, learned policies such as MINERVA incorporate a STOP action to determine whether to continue path expansion \citep{Das2018}. Candidate selection and termination therefore address different stages of the traversal process: one determines which edge to follow, while the other determines whether to continue exploring the graph.

These decisions are also distinct from assessing whether the retrieved path contains sufficient evidence to answer the query. A traversal policy determines when to stop expanding a path, but that stopping decision does not by itself establish that the collected evidence supports an answer. This distinction provides the context for studying retrieval methods that combine query dependent edge selection and traversal control with an explicit assessment of evidence sufficiency before generation.

\section{Problem Formulation}
\label{sec:problem}

Graph-based RAG retrieves evidence
by traversing a knowledge graph. The required evidence path varies
in length across queries. The retriever must select an edge at each
step and determine when to terminate traversal. After path
construction, a separate verification step assesses whether the
retrieved evidence supports answer generation. This paper formulates
graph retrieval as a sequential decision problem with adaptive
traversal and path-sufficiency verification.

\begin{algorithm}[t]
\caption{GRASP Inference Overview}
\label{alg:grasp_overview}
\begin{algorithmic}[1]
\Require Query $q$, graph $G$, entry entity $v_0$,
maximum attempts $I_{\max}$, maximum hops $T_{\max}$
\Ensure Answer $\widehat{A}$ or abstention

\State Encode $q$; initialize $\mathcal{X}(v)\gets\emptyset$
\Comment{Stage A}
\For{$i=1$ to $I_{\max}$}
    \State Retrieve $P_i$ from $v_0$ using $\pi$ and $\pi_{\mathrm{stop}}$,
    excluding $\mathcal{X}$, within $T_{\max}$ hops
    \Comment{Stage B}
    \State Evaluate $S(q,P_i)$ if $P_i\neq\emptyset$
    \Comment{Stage C}
    \If{$P_i\neq\emptyset$ \textbf{and} $S(q,P_i)=1$}
        \Comment{Stage D}
        \State \Return $\operatorname{LLM}(q,\operatorname{Serialize}(P_i))$
        \Comment{Stage E}
    \ElsIf{$P_i\neq\emptyset$}
        \State Let $a^{-}$ be the lowest-margin action in $P_i$
        \State $\mathcal{X}(\operatorname{src}(a^{-}))
        \gets\mathcal{X}(\operatorname{src}(a^{-}))\cup\{a^{-}\}$
    \EndIf
\EndFor
\State \Return abstention
\end{algorithmic}
\end{algorithm}

\paragraph{Problem setup.}
Let $G=(V,E,R)$ denote a directed knowledge graph, where $V$ is the
entity set, $E$ is the edge set, and $R$ is the set of relation types
\citep{Bodenreider2004,Himmelstein2017,JiKGSurvey2022}.
Given a query $q$ and an entry entity $v_0$, the retriever constructs
a path
\begin{equation}
P=(v_0,r_1,v_1,\ldots,r_T,v_T),\,
(v_{t-1},r_t,v_t)\in E,\quad t=1,\ldots,T.
\label{eq:path}
\end{equation}
Here, $T$ denotes the traversal depth and is determined during
inference rather than fixed in advance.

\paragraph{Adaptive traversal.}
At step $t$, a policy $\pi$ selects an outgoing edge
$a_t=(r_t,v_t)$ according to the query and the current traversal
state $s_t$:
\begin{equation}
\hfil a_t\sim\pi(a_t\mid q,s_t).\hfil
\label{eq:policy}
\end{equation}
The stopping gate determines whether traversal terminates:
\begin{equation}
z_t=\pi_{\mathrm{stop}}(q,s_t,t),
\qquad z_t\in\{0,1\},
\label{eq:stop}
\end{equation}
where $z_t=1$ indicates termination. These decisions jointly
determine the retrieved path and its length, following the
sequential traversal formulation used in prior multi-hop reasoning
methods \citep{Xiong2017,Das2018}.

\paragraph{Path-sufficiency verification.}
A completed path may contain relevant entities without providing
all the relations needed to answer the query. We therefore define
a path-sufficiency function
\begin{equation}
\hfil S(q,P)\in\{0,1\}.\hfil
\label{eq:sufficiency}
\end{equation}
where $S(q,P)=1$ indicates that $P$ provides sufficient evidence
for answering $q$. The verifier evaluates the completed path after
traversal. If the path is insufficient, GRASP can initiate another
retrieval attempt, subject to the inference budget. Generation then
uses the accepted path, or follows the predefined budget-exhaustion
policy if verification does not succeed.

\paragraph{Evaluation criteria.}
We evaluate retrieval using task specific accuracy and path recall.
For entity-answer benchmarks, terminal entity accuracy is defined as
\begin{equation}
\mathrm{Acc}_{\mathrm{ent}}(\pi)
=
\frac{1}{|Q|}
\sum_{q\in Q}
\mathbf{1}\!\left[v_T(q)=v^*(q)\right].
\label{eq:acc_ent}
\end{equation}
For paragraph-level answers, we use answer entity recall:
\begin{equation}
\mathrm{Acc}_{\mathrm{par}}(\pi)
=
\frac{1}{|Q|}
\sum_{q\in Q}
\frac{
|E(A^*(q))\cap V(P_\pi(q))|
}{
|E(A^*(q))|
}.
\label{eq:acc_par}
\end{equation}
Here, $A^*(q)$ denotes the reference answer, $E(A^*(q))$ its
annotated salient entities, and $V(P_\pi(q))$ the entities covered
by the retrieved path. These measures are task-specific proxies
and are not directly interchangeable.

We also report path recall, denoted by
$\overline{\mathrm{PathRec}}$, to measure coverage of the annotated
reasoning chain, and the unsupported-claim rate,
$\overline{\mathrm{Unsupported}}$, to measure the proportion of
generated claims that lack support from the retrieved evidence.
The precise definition of the latter is provided in
Section~\ref{sec:method}. We report these metrics separately to
distinguish retrieval accuracy, path coverage, and generation
grounding.

\paragraph{Inference procedure.}
GRASP organizes inference into five stages. Stage~A prepares the
query and initializes retrieval. Stage~B performs adaptive traversal
through candidate edge scoring and stopping. Stage~C evaluates path
sufficiency. Stage~D uses that assessment to accept the evidence or
initiate another retrieval attempt within the specified budget.
Stage~E generates the answer from accepted evidence.
Algorithm~\ref{alg:grasp_overview} summarizes these stages.

\begin{figure*}[t]
    \centering
    \includegraphics[width=\textwidth]{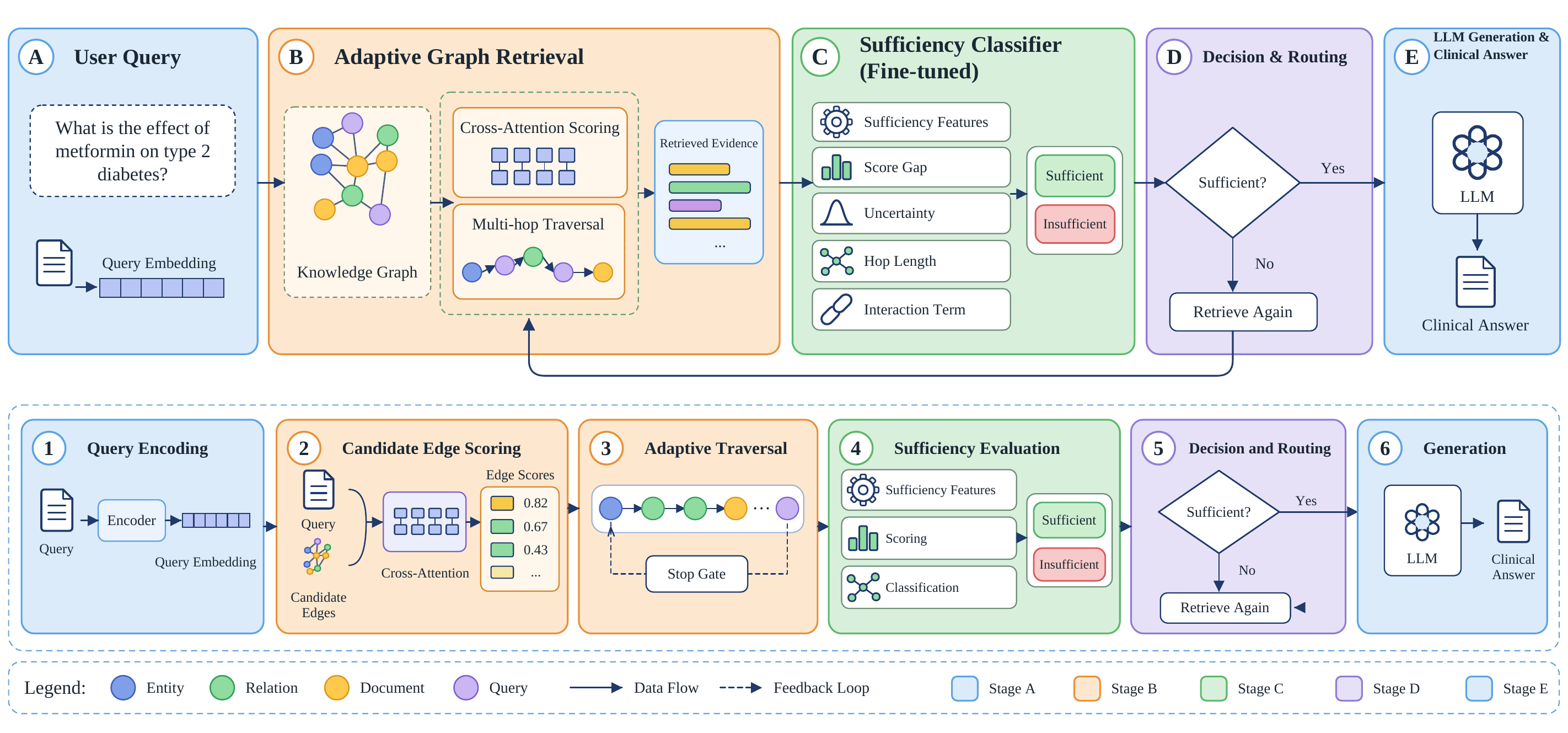}
    \caption{Overview of the proposed GRASP framework. Stage A
    encodes the input query. Stage B performs adaptive graph retrieval
    through cross-attention edge scoring and multi-hop traversal with
    a learned stopping gate. Stage C evaluates the retrieved path
    using path representations and retrieval-derived sufficiency
    features. Stage D determines whether the evidence is sufficient
    for generation or another retrieval attempt is required. Stage E
    generates the clinical answer using the accepted evidence.}
    \label{fig:framework}
\end{figure*}

\section{Methodology}
\label{sec:method}

GRASP is designed to improve graph-based retrieval by addressing
three interdependent decisions: candidate-edge selection, traversal
termination, and retrieved-path sufficiency. As illustrated in
Fig.~\ref{fig:framework}, the framework consists of five stages.
Stage~A encodes the input query and provides its representations to
the retrieval process. Stage~B performs adaptive graph retrieval
through query-conditioned edge scoring and multi-hop traversal.
Stage~C evaluates the structural sufficiency of the retrieved path,
and Stage~D determines whether the path can be accepted or whether
another retrieval attempt is required. Once a path is accepted,
Stage~E uses the retrieved evidence to generate the clinical answer.
Thus, Stages~A and E define the input and output interfaces of GRASP,
while Stages~B, C, and D form the retrieval, verification, and
correction process.

The framework contains two closely connected learning components
within this five-stage process. The retriever in Stage~B combines
query-adaptive edge scoring with a learned stopping mechanism to
construct a query-specific path. The verifier in Stage~C evaluates
the resulting path using its structural representation and retrieval
confidence statistics. Stage~D connects these two components at
inference time: it accepts a path that meets the sufficiency
criterion or initiates another retrieval attempt when the path is
predicted to be insufficient. The accepted path is then passed to
Stage~E for answer generation. The two learning components are
trained separately in two stages and coupled during inference
through the verification and re-retrieval process.


\subsection{Query-Adaptive Edge Scoring}
\label{sec:edge_scoring}

The first component of Stage~B determines which outgoing edge should
be selected at each traversal step. In graph-based retrieval, the
relevance of a candidate edge depends not only on its semantic
similarity to the query as a whole, but also on how its relation and
destination entity correspond to specific query tokens. A single
global similarity score may overlook these fine-grained interactions.
GRASP therefore combines global semantic matching with token-level
cross-attention to obtain a more query-specific assessment of each
candidate action.

Given a query $q$, the query encoder produces token representations
$\mathbf{H}_q\in\mathbb{R}^{L_q\times d}$ and a global representation
$\mathbf{c}_q\in\mathbb{R}^{d}$, where $L_q$ is the query length and
$d$ is the representation dimension. The projected token
representations used for cross-attention are denoted by
$\widetilde{\mathbf{H}}_q$. At traversal step $t$, each available
action $a_k=(r_k,v_k')$, consisting of a relation $r_k$ and a
destination entity $v_k'$, is encoded as
$\mathbf{e}_{a_k}\in\mathbb{R}^{d}$.

First, the global relevance between the query and a candidate action
is measured by cosine similarity:
\begin{equation}
s_{\mathrm{global}}(q,a_k)
=
\frac{\mathbf{c}_q^\top\mathbf{e}_{a_k}}
{\|\mathbf{c}_q\|_2\|\mathbf{e}_{a_k}\|_2}.
\label{eq:global_score}
\end{equation}
This score provides an overall semantic compatibility signal
between the query and the candidate edge. However, global
representations may compress token-specific information and fail
to distinguish candidates that are relevant to different aspects
of the same query.

To capture fine-grained relevance, GRASP introduces a
cross-attention mechanism in which each candidate edge attends to
the query tokens. Specifically, the candidate edge representation
is projected into a query vector, while the projected query tokens
provide the keys and values:
\begin{align}
\boldsymbol{\alpha}_k
&=
\operatorname{softmax}
\left(
\frac{
(\mathbf{W}_Q\mathbf{e}_{a_k})^\top
(\mathbf{W}_K\widetilde{\mathbf{H}}_q)^\top
}{\sqrt{d_a}}
\right),\\
\mathbf{c}_{a_k}
&=
\boldsymbol{\alpha}_k
\mathbf{W}_V\widetilde{\mathbf{H}}_q,\\
s_{\mathrm{token}}(q,a_k)
&=
\operatorname{sim}
\left(
\mathbf{W}_O[\mathbf{e}_{a_k};\mathbf{c}_{a_k}],
\mathbf{c}_q
\right),
\label{eq:token_score}
\end{align}
where $\mathbf{W}_Q$, $\mathbf{W}_K$, $\mathbf{W}_V$, and
$\mathbf{W}_O$ are learnable projections, $d_a$ is the attention
key dimension, and $\operatorname{sim}$ denotes the similarity
function used by the scorer. The attention weights
$\boldsymbol{\alpha}_k$ determine the relative contribution of
individual query tokens to the representation associated with
candidate $a_k$. Consequently, different candidate edges can
extract different query-specific contextual information.

The global and token-level scores are then combined through a
weighted fusion:
\begin{equation}
s_k^{(t)}
=
\gamma s_{\mathrm{global}}(q,a_k)
+
(1-\gamma)s_{\mathrm{token}}(q,a_k),
\quad \gamma\in[0,1].
\label{eq:combined_score}
\end{equation}
The coefficient $\gamma$ controls the relative contribution of
global semantic compatibility and fine-grained token-level
alignment. This combination allows the retriever to consider both
the overall relevance of an edge to the query and its correspondence
to specific query components. At each traversal step, the action
with the highest combined score is selected from the available
candidates.

\subsection{Adaptive Graph Traversal and Termination}
\label{sec:adaptive_stopping}

Selecting relevant individual edges does not by itself determine
whether the resulting path is appropriate for a given question.
Questions differ in the amount of graph evidence required to reach
a useful answer: some can be addressed through a short path, whereas
others require multiple relational steps. A fixed traversal depth
may therefore introduce unnecessary edges for simpler questions
or terminate before sufficient evidence has been collected for
more complex ones. GRASP addresses this issue through a learned
stopping gate that adapts traversal termination to the current
retrieval state.

After an action is selected, its representation is incorporated
into the recurrent traversal state. The state is subsequently
refined by attending over the action history, allowing the retriever
to retain information about previously selected edges rather than
making each decision independently. This history-aware state
provides a representation of the current traversal context and
serves as an input to the stopping gate. In this way, termination
decisions can reflect both the query and the evidence accumulated
along the current path.

Let $\widehat{\mathbf{s}}_t$ denote the refined traversal state
after selecting the $t$-th action. The stopping probability is
computed as
\begin{equation}
p_{\mathrm{stop}}^{(t)}
=
\sigma\left(
g_\theta(\mathbf{c}_q,\widehat{\mathbf{s}}_t,t)
\right),
\label{eq:stop_probability}
\end{equation}
where $g_\theta$ is a learned scoring function, $\sigma$ is the
sigmoid function, and $t$ represents the current hop depth. The
stopping gate jointly considers the global query representation,
the refined traversal state, and the number of actions already
selected. Traversal terminates when
$p_{\mathrm{stop}}^{(t)}>\tau_{\mathrm{stop}}$ or when the maximum
hop count $T_{\max}$ is reached.

The combination of query information and traversal context is
important because the stopping decision should not depend on
retrieval depth alone. A path may reach a meaningful state after
relatively few hops, while another query may require further
exploration. The learned gate provides a query- and state-dependent
termination decision, while $T_{\max}$ imposes an upper bound on
the traversal length. If no action remains available at the current
entity, the attempt ends as a dead end and its incomplete prefix
is not passed to the sufficiency verifier.

The edge scorer and stopping gate are trained jointly in Stage~B.
Their training objective combines positive-edge discrimination,
contrastive learning, pairwise ranking, and stopping supervision:
\begin{equation}
\mathcal{L}_{B}
=
\lambda_{\mathrm{NCE}}\mathcal{L}_{\mathrm{NCE}}
+
\lambda_{\mathrm{con}}\mathcal{L}_{\mathrm{con}}
+
\lambda_{\mathrm{rank}}\mathcal{L}_{\mathrm{rank}}
+
\lambda_{\mathrm{stop}}\mathcal{L}_{\mathrm{stop}},
\label{eq:stage_b_loss}
\end{equation}
where $\mathcal{L}_{\mathrm{NCE}}$ supervises positive-edge
discrimination, $\mathcal{L}_{\mathrm{con}}$ provides contrastive
supervision, and $\mathcal{L}_{\mathrm{rank}}$ encourages the
retriever to rank relevant actions above competing candidates.
The stopping term $\mathcal{L}_{\mathrm{stop}}$ is a binary
cross-entropy loss for traversal termination. The coefficients
$\lambda_{\mathrm{NCE}}$, $\lambda_{\mathrm{con}}$,
$\lambda_{\mathrm{rank}}$, and $\lambda_{\mathrm{stop}}$ control
the relative contributions of the four objectives. The stopping
loss weight follows the warm-up schedule used during training.

\subsection{Path-Sufficiency Verification}
\label{sec:path_verification}

Although adaptive edge selection and traversal termination help
construct query-relevant paths, high edge scores do not necessarily
indicate that the assembled path contains all the evidence required
to support an answer. A path may include relevant entities while
omitting an important relation or an intermediate connection.
GRASP therefore introduces a path-sufficiency verifier in Stage~C,
as illustrated in Fig.~\ref{fig:framework}. Instead of treating
retriever confidence as a direct measure of evidence sufficiency,
the verifier evaluates the assembled path using both its sequential
representation and statistics derived from the retrieval process.

For a retrieved path $P=(a_1,\ldots,a_T)$, a bidirectional GRU
encodes the ordered action sequence into a path representation
$\mathbf{h}_P$. The sequential encoder preserves the order of
the selected actions, providing a representation of the retrieved
path rather than an unordered collection of candidate edges.

In addition to this path representation, the verifier uses six
retrieval confidence statistics: mean edge score
$\overline{s}$, mean hop margin $\overline{\delta}$, mean
candidate entropy $\overline{H}$, maximum edge score $s_{\max}$,
minimum hop margin $\delta_{\min}$, and path length $T$. Each
hop margin $\delta_t$ is defined as the score difference between
the selected action and its strongest competing candidate at
the corresponding traversal step. These statistics capture
different aspects of the retrieval process. The mean edge score
summarizes the overall scoring level, whereas the maximum edge
score records the strongest local retrieval signal. The mean
and minimum hop margins characterize the separation between
selected actions and their alternatives, providing information
about average and weakest-step selection confidence. Mean
candidate entropy reflects uncertainty in candidate selection,
while path length provides information about the extent of
the retrieved traversal.

The path representation and retrieval statistics are concatenated
to form the verifier input:
\begin{equation}
\mathbf{f}(q,P)
=
[
\mathbf{h}_P;
\overline{s};
\overline{\delta};
\overline{H};
s_{\max};
\delta_{\min};
T
].
\label{eq:verifier_features}
\end{equation}
Scalar features are standardized using statistics computed from
the Stage~C training set. Combining the path representation with
retrieval-derived statistics allows the verifier to consider both
the content and structure of the retrieved path and the confidence
characteristics of the decisions that produced it.

An MLP maps the resulting feature vector to the probability that
the path is insufficient:
\begin{equation}
p_{\mathrm{ins}}
=
\sigma\left(
h_\phi(\mathbf{f}(q,P))
\right),
\label{eq:insufficiency_probability}
\end{equation}
where $h_\phi$ denotes the verifier network and
$p_{\mathrm{ins}}$ is the predicted insufficiency probability.
The verifier is trained in Stage~C using binary cross-entropy:
\begin{equation}
\mathcal{L}_{C}
=
-\frac{1}{N}
\sum_{i=1}^{N}
\left[
y_i\log p_i+
(1-y_i)\log(1-p_i)
\right],
\label{eq:stage_c_loss}
\end{equation}
where $N$ is the number of training examples, $y_i=1$ denotes
an insufficient path, and $p_i$ is its predicted insufficiency
probability. The training labels are derived from the structural
path-quality criteria and constructed perturbations described
in the experimental protocol. These labels provide proxy
supervision for evidence sufficiency rather than direct guarantees
of semantic correctness.

\begin{algorithm*}[t]
\caption{GRASP: Query-Adaptive Graph Retrieval with Path Sufficiency Verification}
\label{alg:grasp_inference}
\begin{algorithmic}[1]
\Require Query $q$, knowledge graph $\mathcal{G}$, entry entity $v_0$,
trained retrieval model $\mathcal{M}_B$, trained sufficiency classifier
$\mathcal{M}_C$, stopping threshold $\tau_{\mathrm{stop}}$,
sufficiency threshold $\tau_{\mathrm{cls}}$, maximum hop count
$T_{\max}$, maximum retrieval attempts $I_{\max}$
\Ensure Generated answer $\widehat{A}$ or abstention

\State Encode $q$ to obtain global representation $\mathbf{c}_q$
and token representations $\widetilde{\mathbf{H}}_q$
\State Initialize exclusion sets $\mathcal{X}(v)\gets\emptyset$
for all entities $v\in\mathcal{V}$

\For{$i=1$ to $I_{\max}$}
    \State Initialize path $P_i\gets\emptyset$, current entity
    $v\gets v_0$, and recurrent state $\mathbf{s}_0\gets\mathbf{c}_q$
    \State Initialize action history $\mathbf{E}_0\gets\emptyset$

    \For{$t=1$ to $T_{\max}$}
        \State Construct available actions
        $\mathcal{A}_t\gets
        \{(r,v'):(v,r,v')\in\mathcal{E}\}
        \setminus\mathcal{X}(v)$
        \If{$\mathcal{A}_t=\emptyset$}
            \State \textbf{break}
        \EndIf

        \State Score each $a_k\in\mathcal{A}_t$ using
        Eqs.~\eqref{eq:global_score}--\eqref{eq:combined_score},
        conditioned on $\mathbf{c}_q$,
        $\widetilde{\mathbf{H}}_q$, and the refined state
        from the preceding hop
        \State Select $a_t^*\gets\arg\max_{a_k\in\mathcal{A}_t}s_k^{(t)}$
        \State Append $a_t^*$ to $P_i$
        \State Update recurrent state:
        $\mathbf{s}_t\gets
        \operatorname{GRU}(\mathbf{e}_{a_t^*},\mathbf{s}_{t-1})$
        \State Append $\mathbf{e}_{a_t^*}$ to action history
        $\mathbf{E}_t$
        \State Refine state:
        $\widehat{\mathbf{s}}_t\gets
        \operatorname{LayerNorm}\bigl(
        \mathbf{s}_t+
        \operatorname{Attention}(\mathbf{s}_t,\mathbf{E}_t,\mathbf{E}_t)
        \bigr)$
        \State Compute stopping probability
        $p_{\mathrm{stop}}^{(t)}\gets\sigma(g_t)$ using
        Eq.~\eqref{eq:stop_probability}

        \If{$p_{\mathrm{stop}}^{(t)}>\tau_{\mathrm{stop}}$}
            \State \textbf{break}
        \EndIf
        \State $v\gets\operatorname{dst}(a_t^*)$
    \EndFor

    \If{$P_i=\emptyset$}
        \State \textbf{continue}
    \EndIf

    \State Compute path representation $\mathbf{h}_{P_i}$ and
    confidence features $\mathbf{f}(q,P_i)$
    \State Standardize features using the Stage C training statistics
    \State Compute insufficiency probability
    $p_{\mathrm{ins}}\gets\mathcal{M}_C(q,P_i)$

    \If{$p_{\mathrm{ins}}<\tau_{\mathrm{cls}}$}
        \State $\widehat{A}\gets
        \operatorname{LLM}(q,\operatorname{Serialize}(P_i))$
        \State \Return $\widehat{A}$
    \EndIf

    \State Compute hop margins $\delta_t$ for $P_i$
    \State $t^*\gets\arg\min_{1\leq t\leq |P_i|}\delta_t$
    \State Let $v_{\mathrm{src}}$ be the source entity of action
    $a_{t^*}^*$ in $P_i$
    \State $\mathcal{X}(v_{\mathrm{src}})
    \gets\mathcal{X}(v_{\mathrm{src}})\cup\{a_{t^*}^*\}$
\EndFor

\State \Return abstention
\end{algorithmic}
\end{algorithm*}

\subsection{Verification-Driven Re-Retrieval}
\label{sec:reretrieval}

The output of the sufficiency classifier determines the subsequent
decision and routing process in Stage~D of Fig.~\ref{fig:framework}.
A retrieved path is accepted when
$p_{\mathrm{ins}}<\tau_{\mathrm{cls}}$, in which case it is passed
to the LLM for answer generation. If
$p_{\mathrm{ins}}\geq\tau_{\mathrm{cls}}$, the path is treated as
insufficient and another retrieval attempt is initiated when the
remaining retrieval budget permits. This verification step
distinguishes GRASP from a retrieval process that directly passes
the highest-scoring path to generation without an explicit
path-level assessment.

To encourage exploration of an alternative path, GRASP identifies
the traversal step with the smallest hop margin:
\begin{equation}
t^*=\arg\min_{t\in\{1,\ldots,T\}}\delta_t.
\label{eq:min_margin_step}
\end{equation}
The selected action at this step is excluded from the candidate
actions available at its source entity. The next retrieval attempt
then restarts from the original entry entity. Exclusion sets are
maintained across attempts, preventing the same excluded action
from being selected again at the corresponding source entity.

The re-retrieval procedure is designed as a confidence-guided
heuristic for exploring alternative paths. By modifying a
low-margin decision, it directs subsequent retrieval away from
one of the least confident local choices in the previous attempt.
However, a low-margin action is not necessarily incorrect, and
changing a single action does not guarantee recovery from every
retrieval failure. In particular, the procedure may not resolve
errors that arise from an earlier traversal decision or from
premature termination.

The total number of retrieval attempts, including the initial
attempt, is bounded by $I_{\max}$. If a sufficient path is found
within this budget, it is passed to answer generation. Otherwise,
the system follows the predefined budget-exhaustion policy.
The complete interaction between adaptive traversal, path
verification, and re-retrieval is summarized in
Algorithm~\ref{alg:grasp_inference}.

\subsection{Two-Stage Training}
\label{sec:training_inference}

GRASP adopts a two-stage training strategy to learn retrieval
and path verification separately. In Stage~B, the candidate-edge
scorer and stopping gate are jointly optimized using the
retrieval-related objectives described in
Section~\ref{sec:adaptive_stopping}. After the retriever converges,
its retrieval outputs and confidence statistics are used to
construct path-level features for Stage~C. The sufficiency
classifier is then trained using these features and the
path-quality proxy labels described in the experimental protocol.

The two stages do not share a joint end-to-end training objective.
Instead, Stage~C learns to assess paths produced by the trained
Stage~B retriever. This separation allows the verifier to use
retrieval-specific confidence information while maintaining a
distinct training objective for path-level sufficiency assessment.
During inference, the two trained components are coupled through
the verification-driven re-retrieval process. Only a path accepted
by the verifier is passed to the generation component, as shown
in Figure ~\ref{fig:framework}.

\section{Experiments and Results}
\label{sec:experiments}
\providecommand{\mpb}{MINERVA-PubMed\allowbreak-BERT}

This section tests whether the three design choices of GRASP improve
retrieval and generation: token-level candidate scoring with adaptive
stopping (Stage~B), sufficiency verification (Stage~C), and local
re-retrieval (Stage~D). We first compare GRASP with the baselines, then
isolate the contribution of each component through a controlled
comparison and ablation studies, and finally examine the reliability of
the verification stages and the sensitivity of the loop.

\subsection{Experimental Protocol}
\label{sec:experimental_protocol}

We evaluate GRASP on three biomedical question-answering benchmarks
with different graph structures and annotation properties. The
protocol covers dataset preparation, model training, structural
sufficiency supervision, and generation evaluation.

\paragraph{Datasets and data splits.}

We use RiTeK, MedKGent-KG, and a PMC-based biomedical corpus
\citep{LiRiTeK2024,MedKGent2024,IslamajPMC2021}. Their main
characteristics and data splits are summarized in
Table~\ref{tab:dataset_protocol}.

\begin{table*}[t]
\centering
\small
\setlength{\tabcolsep}{7pt}
\renewcommand{\arraystretch}{1.15}
\caption{Dataset characteristics and experimental data splits.}
\label{tab:dataset_protocol}
\begin{tabular}{lccccc}
\toprule
Dataset & Graph scale & Supervision & Training & Development & Test \\
\midrule
RiTeK
& $\sim$3K entities
& Query-level paths
& 3,500 & 1,500 & 100 \\

MedKGent-KG
& 156K entities, 2.97M triples
& Expert-scored triples
& 672 & 288 & 100 \\

PMC
& Query-specific graphs
& Sentence-level distant supervision
& 3,500 & 1,500 & 100 \\
\bottomrule
\end{tabular}
\end{table*}

RiTeK provides annotated reasoning paths for evaluating terminal
entity retrieval and intermediate evidence recovery. MedKGent-KG
provides a larger and denser biomedical graph, together with
1,060 expert-scored triple records. These annotations are used
as query-associated reference evidence rather than being treated
as exhaustive graph-level or ordered path annotations. The ratio
of triples to entities is approximately 19.04, indicating the
larger candidate space considered in this benchmark.

PMC does not provide a fixed knowledge graph or annotated
reasoning paths. We therefore construct a query-specific graph
for each question by retrieving relevant full-text articles,
extracting biomedical entities and relations from their sentences,
and assembling the extracted relations into a directed graph.
Entity normalization is applied before graph assembly, and
extracted edges retain their source-sentence information for
evidence inspection. Reference answers are excluded from
test-time graph construction. Since PMC lacks annotated
reasoning paths, Path Recall is not reported for this dataset.

RiTeK and PMC each contribute 5,000 instances for training
and development, whereas MedKGent-KG contributes its
1,060 available expert-annotated records. The training and
development portions follow a 70:30 split, with an additional
100 held-out test instances per benchmark. Splitting is performed
before generating path perturbations. For PMC, questions
associated with the same source article are kept within the
same partition. The held-out test sets are excluded from
training, checkpoint selection, and threshold optimization.
For MedKGent-KG, target annotations are separated before
training so that the target evidence of a held-out query is
not exposed to Stage~B.

\paragraph{Baselines and implementation.}

We compare GRASP with five baselines: BFS-2, BFS-3,
G-Retriever, GNN-RAG, and MINERVA
\citep{He2024,Mavromatis2024,Das2018}. BFS-2 and
BFS-3 use fixed traversal depths of two and three hops,
respectively. G-Retriever and GNN-RAG represent graph-based
retrieval for language model reasoning, while MINERVA
provides a learned multi-hop traversal baseline. All methods
use the same benchmark queries, graph resources, and
answer-generation model. A separate controlled comparison
uses MINERVA equipped with PubMedBERT to distinguish
the effect of the encoder backbone from that of the GRASP
architecture.

GRASP uses PubMedBERT as its shared textual encoder
\citep{Gu2021}. Stage~B jointly trains the candidate-edge
scorer and stopping gate, while Stage~C independently
trains the sufficiency classifier. Both stages use AdamW
optimization \citep{Loshchilov2019}. The pretrained encoder
uses a lower learning rate than the newly initialized layers,
with linear warm-up followed by cosine decay. The principal
inference settings are $I_{\max}=3$ total retrieval attempts,
including the initial attempt, and
$\tau_{\mathrm{cls}}=0.50$. The stopping threshold and
maximum hop count are selected using development data.

All experiments use an NVIDIA GeForce RTX~3080 GPU
with 16\,GB system memory. Gemini-2.5 Flash is used
as the common answer generator without task-specific
fine-tuning. The same prompt structure and decoding
configuration are applied across the compared retrieval
methods.

\paragraph{Sufficiency supervision and generation evaluation.}

Stage~C is trained using structural proxy labels rather than
generation-level Unsupported Claim Rate, which is available
only after answer generation. Complete reference evidence
is labelled sufficient ($y=0$), while evidence that fails
to cover the required structural information is labelled
insufficient ($y=1$). RiTeK provides ordered reference
paths, whereas MedKGent-KG provides query-associated
annotated evidence triples. The latter are used to assess
evidence coverage and are not treated as complete ordered
gold paths. PMC is excluded from supervised gold-path
construction because it has no corresponding path annotations.

Insufficient examples are generated through six perturbations:
75\% truncation, 50\% truncation, node substitution, relation
swap, over-extension, and reverse-path corruption. Truncation
removes required evidence from the path, while node and
relation substitutions alter intermediate retrieval decisions.
Over-extension introduces an irrelevant or conflicting
continuation, and reverse-path corruption changes the direction
or order of required relations. Additional hard negatives
are obtained from paths generated by Stage~B. Perturbed
examples derived from the same source query or path remain
within the same data partition to prevent source-level leakage.
These structural labels serve as proxies for retrieval
sufficiency and are not treated as guarantees of factual
correctness.

Generated answers are decomposed into atomic claims and
evaluated against the retrieved path using natural language
inference. The
\texttt{microsoft/deberta-xlarge-mnli} checkpoint is used
for the NLI assessment and BERTScore
\citep{ZhangBERT2020}. Each claim is classified as entailed,
neutral, or contradicted by the retrieved evidence. Only
entailed claims are considered supported. Neutral and
contradicted claims contribute to the Unsupported Claim
Rate. Accordingly, the Path-Supported Atomic Fact Rate
(PSAFR) is the proportion of generated atomic claims
entailed by the retrieved path, while the Path-Neutral
Rate (PNR) and Path-Contradiction Rate (PCR) denote
the proportions of neutral and contradicted claims,
respectively. We therefore obtain

\begin{equation}
\mathrm{UCR}
=
\mathrm{PNR}+\mathrm{PCR}
=
1-\mathrm{PSAFR}.
\end{equation}

A neutral claim is not necessarily factually incorrect;
it indicates that the retrieved path does not provide
sufficient support for that claim.

\paragraph{Metrics and statistical analysis.}

Retrieval performance is measured using Hits@1, Hits@3,
MRR, and Path Recall. Path Recall measures the proportion
of query-associated reference triples covered by the
retrieved path. It is a set-based coverage measure rather
than an exact ordered-path match. For PMC, the
paragraph-level answer entity recall defined in
Eq.~\eqref{eq:acc_par} is used as a separate retrieval
proxy and is not directly compared with terminal-entity
Hits@1 on the other benchmarks.

Generation quality is evaluated using Token F1 and
BERTScore, while evidence groundedness is measured by
UCR. Higher values are preferable for retrieval and
generation quality metrics, whereas lower UCR indicates
better evidence support. Absolute differences between
percentage-valued metrics are reported in percentage
points (pp).

Each trainable configuration is evaluated using five
independent random seeds, and retrieval results are
reported as mean $\pm$ standard deviation. For the
BFS baselines, the seeds control tie-breaking among
equally ranked candidate edges rather than model
initialization. Generation results are first aggregated
over repeated decoder runs within each question and
then summarized across questions. Paired bootstrap
tests use the same held-out questions for both systems,
with 10,000 question-level resamples. This separates
training and decoding variability from uncertainty
associated with the test queries.

\subsection{Main Results}
\label{sec:main_results}

\begin{figure*}[!t]
\centering
\includegraphics[width=0.95\textwidth]{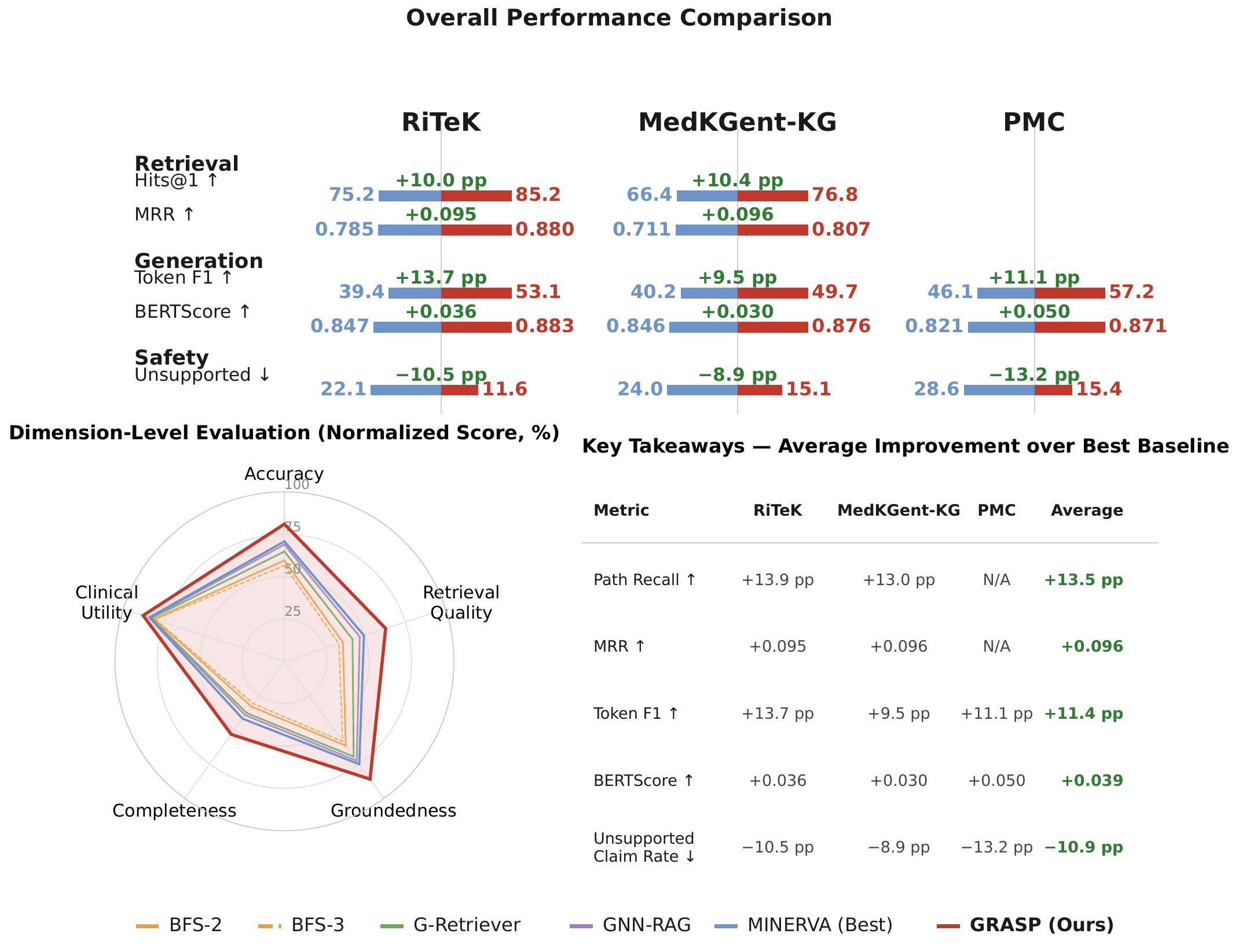}
\caption{Comparison between GRASP and the strongest baseline on each
dataset. The upper block reports retrieval, generation, and safety
metrics; the best baseline is the highest-scoring system among BFS-2,
BFS-3, G-Retriever, GNN-RAG, and MINERVA on that dataset. The lower
left shows dimension-level normalized scores averaged over the three
datasets, and the lower right lists the average improvement per metric.}
\label{fig:overview}
\end{figure*}

Figure~\ref{fig:overview} compares GRASP with the strongest baseline on
every dataset. GRASP leads on all reported metrics, and the advantage
covers accuracy, retrieval quality, completeness, and groundedness
rather than terminal accuracy alone.

Table~\ref{tab:main_retrieval} reports the retrieval results. GRASP
improves Hits@1 over MINERVA by 10.0\% on RiTeK and 10.4\% on
MedKGent-KG, and Path Recall by 13.9\% and 13.0\%. The Path Recall gain
exceeds the Hits@1 gain on both datasets, which indicates that
token-level scoring with adaptive stopping recovers the intermediate
relations of the annotated chain and not only the final entity. The
gains are similar on the two datasets, and we did not test whether they
differ.

\begin{table*}[!t]
\centering
\small
\setlength{\tabcolsep}{5pt}
\renewcommand{\arraystretch}{1.05}
\caption{Main retrieval results (mean $\pm$ standard deviation over five
independently trained runs). Path Recall is not reported for PMC, which
lacks gold reasoning paths.}
\label{tab:main_retrieval}
\begin{tabular}{lcccccccc}
\toprule
\multirow{2}{*}{Model}
& \multicolumn{4}{c}{RiTeK}
& \multicolumn{4}{c}{MedKGent-KG} \\
\cmidrule(lr){2-5}
\cmidrule(lr){6-9}
& Hits@1 & Hits@3 & MRR & Path Rec.
& Hits@1 & Hits@3 & MRR & Path Rec. \\
\midrule
BFS-2
& $63.2\pm1.2$ & $74.1\pm1.1$ & $0.672\pm0.007$ & $40.5\pm1.6$
& $55.9\pm1.1$ & $67.4\pm1.0$ & $0.601\pm0.007$ & $32.1\pm1.4$ \\
BFS-3
& $60.7\pm1.3$ & $71.8\pm1.2$ & $0.648\pm0.008$ & $37.9\pm1.5$
& $52.4\pm1.2$ & $64.8\pm1.1$ & $0.573\pm0.008$ & $29.6\pm1.5$ \\
G-Retriever
& $68.4\pm1.1$ & $79.2\pm1.0$ & $0.722\pm0.007$ & $46.1\pm1.5$
& $61.3\pm1.0$ & $73.9\pm1.0$ & $0.665\pm0.006$ & $38.4\pm1.4$ \\
GNN-RAG
& $72.1\pm1.0$ & $82.6\pm0.9$ & $0.756\pm0.006$ & $50.3\pm1.4$
& $65.8\pm1.0$ & $77.2\pm0.9$ & $0.705\pm0.006$ & $43.2\pm1.3$ \\
MINERVA
& $75.2\pm0.8$ & $84.7\pm0.8$ & $0.785\pm0.005$ & $53.8\pm0.9$
& $66.4\pm1.1$ & $77.8\pm0.9$ & $0.711\pm0.007$ & $44.9\pm1.3$ \\
GRASP
& $\mathbf{85.2\pm0.8}$ & $\mathbf{92.0\pm0.6}$ & $\mathbf{0.880\pm0.004}$ & $\mathbf{67.7\pm0.9}$
& $\mathbf{76.8\pm0.8}$ & $\mathbf{86.2\pm0.8}$ & $\mathbf{0.807\pm0.006}$ & $\mathbf{57.9\pm1.0}$ \\
\bottomrule
\end{tabular}
\end{table*}

On MedKGent-KG, held-out triples are separated before path construction
and Stage~B never sees the target triple during training. A blind-tail
evaluation (Section~\ref{sec:blind_tail}) further tests whether the
query template leaks the answer.

\subsection{Generation Results and Blind-Tail Evaluation}
\label{sec:generation_results}

Table~\ref{tab:main_generation} reports generation quality.
Unsupported Claim Rate is the fraction of generated claims that are not
entailed by the retrieved path. It counts claims that are neutral to the
path as well as claims that contradict it, so it is not a factual-error
rate.

\begin{table*}[!t]
\centering
\small
\setlength{\tabcolsep}{4pt}
\renewcommand{\arraystretch}{1.05}
\caption{Generation results. Statistics are aggregated per question
before the mean and standard deviation are computed. Unsup.\ denotes
Unsupported Claim Rate.}
\label{tab:main_generation}
\resizebox{\textwidth}{!}{%
\begin{tabular}{lcccccccccc}
\toprule
\multirow{2}{*}{Model}
& \multicolumn{3}{c}{RiTeK}
& \multicolumn{3}{c}{MedKGent-KG}
& \multicolumn{3}{c}{PMC} \\
\cmidrule(lr){2-4}
\cmidrule(lr){5-7}
\cmidrule(lr){8-10}
& Tok. F1 & BERTScore & Unsup. $\downarrow$
& Tok. F1 & BERTScore & Unsup. $\downarrow$
& Tok. F1 & BERTScore & Unsup. $\downarrow$ \\
\midrule
BFS-2
& $31.4\pm1.2$ & $0.821\pm0.005$ & $34.2\pm1.5$
& $29.3\pm1.1$ & $0.812\pm0.005$ & $38.7\pm1.4$
& $38.4\pm1.0$ & $0.793\pm0.004$ & $42.3\pm1.6$ \\
BFS-3
& $29.8\pm1.1$ & $0.813\pm0.005$ & $37.6\pm1.6$
& $27.1\pm1.0$ & $0.804\pm0.005$ & $41.2\pm1.5$
& $36.2\pm1.0$ & $0.788\pm0.004$ & $45.8\pm1.7$ \\
G-Retriever
& $36.9\pm1.2$ & $0.839\pm0.004$ & $27.3\pm1.4$
& $34.7\pm1.1$ & $0.828\pm0.004$ & $30.1\pm1.3$
& $41.7\pm1.0$ & $0.808\pm0.004$ & $34.1\pm1.5$ \\
GNN-RAG
& $37.1\pm1.2$ & $0.843\pm0.004$ & $24.8\pm1.3$
& $36.9\pm1.1$ & $0.835\pm0.004$ & $26.3\pm1.2$
& $43.9\pm1.0$ & $0.814\pm0.004$ & $31.2\pm1.4$ \\
MINERVA
& $39.4\pm1.1$ & $0.847\pm0.004$ & $22.1\pm1.2$
& $40.2\pm1.0$ & $0.846\pm0.004$ & $24.0\pm1.2$
& $46.1\pm0.9$ & $0.821\pm0.004$ & $28.6\pm1.4$ \\
GRASP
& $\mathbf{53.1\pm0.9}$ & $\mathbf{0.883\pm0.003}$ & $\mathbf{11.6\pm0.8}$
& $\mathbf{49.7\pm0.9}$ & $\mathbf{0.876\pm0.003}$ & $\mathbf{15.1\pm0.8}$
& $\mathbf{57.2\pm0.8}$ & $\mathbf{0.871\pm0.003}$ & $\mathbf{15.4\pm0.9}$ \\
\bottomrule
\end{tabular}}
\end{table*}

Relative to MINERVA, GRASP lowers the Unsupported Claim Rate by
10.5 percentage points on RiTeK, 8.9 on MedKGent-KG, and 13.2 on PMC. The largest
reduction occurs on PMC, where structural supervision is weakest. This
agrees with the role of Stage~C: checking the retrieved path matters
most when generated answers contain more claims that can exceed the
retrieved evidence. Paired bootstrap tests against MINERVA
(Table~\ref{tab:significance}) give $p<0.001$ for every comparison. They
establish the complete-system advantage, while component-level effects
are examined in the ablation studies.

\begin{table}[!t]
\centering
\small
\setlength{\tabcolsep}{6pt}
\renewcommand{\arraystretch}{1.05}
\caption{Paired bootstrap tests comparing GRASP with MINERVA. Resampling
is performed over questions.}
\label{tab:significance}
\begin{tabular}{lccc}
\toprule
Dataset & Metric & $\Delta$ & $p$ \\
\midrule
RiTeK
& Hits@1 & $+10.0\%$ & $<0.001$ \\
& Path Recall & $+13.9\%$ & $<0.001$ \\
& Token F1 & $+13.7\%$ & $<0.001$ \\
& Unsupported & $-10.5\%$ & $<0.001$ \\
\midrule
MedKGent-KG
& Hits@1 & $+10.4\%$ & $<0.001$ \\
& Path Recall & $+13.0\%$ & $<0.001$ \\
& Token F1 & $+9.5\%$ & $<0.001$ \\
& Unsupported & $-8.9\%$ & $<0.001$ \\
\midrule
PMC
& Hits@1$^\ast$ & $+12.7\%$ & $<0.001$ \\
& Token F1 & $+11.1\%$ & $<0.001$ \\
& Unsupported & $-13.2\%$ & $<0.001$ \\
\bottomrule
\end{tabular}
\vspace{4pt}

\raggedright\footnotesize
$^\ast$PMC has no annotated reasoning path. Hits@1 here is the
entity-level accuracy $\mathrm{Acc}_{\mathrm{par}}$ of Eq.~(6), the
overlap between the entities on the retrieved path and the salient
entities of the reference answer. It differs from the terminal-entity
accuracy $\mathrm{Acc}_{\mathrm{ent}}$ used for RiTeK and MedKGent-KG
and should not be compared with it directly.
\end{table}

\subsubsection{Blind-Tail Evaluation}
\label{sec:blind_tail}

The MedKGent-KG query template contains the target entity, which could
serve as a shortcut for terminal entity prediction. In the blind-tail
evaluation, the tail entity is removed from the query while the
semantic relation being asked is preserved.

\begin{table*}[!t]
\centering
\small
\setlength{\tabcolsep}{6pt}
\renewcommand{\arraystretch}{1.05}
\caption{Blind-tail evaluation on MedKGent-KG. The original query
contains the annotated tail entity, and the blind-tail query removes it.
The upper block reports Hits@1 and the lower block Path Recall.}
\label{tab:blind_tail}
\begin{tabular}{llccc}
\toprule
Model & Metric & Original & Blind-Tail & Drop \\
\midrule
\mpb & Hits@1 & 68.7 & 49.6 & $-19.1$ \\
GRASP & Hits@1 & \textbf{76.8} & \textbf{59.7} & $-17.1$ \\
\midrule
\mpb & Path Recall & 47.3 & 43.9 & $-3.4$ \\
GRASP & Path Recall & \textbf{57.9} & \textbf{55.8} & $-2.1$ \\
\bottomrule
\end{tabular}
\end{table*}

Masking the tail entity lowers Hits@1 for both models, which confirms
that the original query offers a partial shortcut. The drop is smaller
for GRASP (17.1\% versus 19.1\%). Path Recall does not depend on the
tail entity appearing in the query, and it falls by only 2.1\% for GRASP
and 3.4\% for \mpb, with GRASP keeping its lead. The retrieval advantage
of GRASP is therefore not explained by copying the target entity from
the input. The experiment does not rule out every effect of query
construction.

\subsection{Controlled Comparison: Backbone versus Architecture}
\label{sec:controlled_architecture}

The main comparison cannot separate the retrieval architecture of GRASP
from the switch to a PubMedBERT backbone. Table~\ref{tab:controlled}
therefore compares the original MINERVA, \mpb, three GRASP variants that
each lack one Stage~B component, and the full model, with the training
data and retrieval task held fixed.

\begin{table*}[!t]
\centering
\small
\setlength{\tabcolsep}{5pt}
\renewcommand{\arraystretch}{1.05}
\caption{Controlled architecture comparison on MedKGent-KG (mean
$\pm$ standard deviation over independent training runs). The
controlled interpretation is restricted to the MINERVA family.}
\label{tab:controlled}
\begin{tabular}{lcccc}
\toprule
Variant & Hits@1 & Path Rec. & Tok. F1 & Unsup. $\downarrow$ \\
\midrule
Original MINERVA
& $66.4\pm1.1$ & $44.9\pm1.3$
& $40.2\pm1.0$ & $24.0\pm1.2$ \\
\mpb
& $68.7\pm1.0$ & $47.3\pm1.3$
& $40.8\pm1.0$ & $23.6\pm1.1$ \\
GRASP w/o CA
& $72.9\pm0.9$ & $51.8\pm1.4$
& $44.1\pm1.0$ & $20.7\pm1.0$ \\
GRASP w/o HM
& $75.1\pm0.9$ & $54.0\pm1.3$
& $46.3\pm1.0$ & $18.0\pm0.9$ \\
GRASP w/o Stop
& $72.7\pm0.9$ & $51.2\pm1.4$
& $44.0\pm1.1$ & $19.3\pm1.0$ \\
GRASP
& $\mathbf{76.8\pm0.8}$ & $\mathbf{57.9\pm1.0}$
& $\mathbf{49.7\pm0.9}$ & $\mathbf{15.1\pm0.8}$ \\
\bottomrule
\end{tabular}
\end{table*}

\begin{figure*}[!t]
\centering
\includegraphics[width=0.8\linewidth]{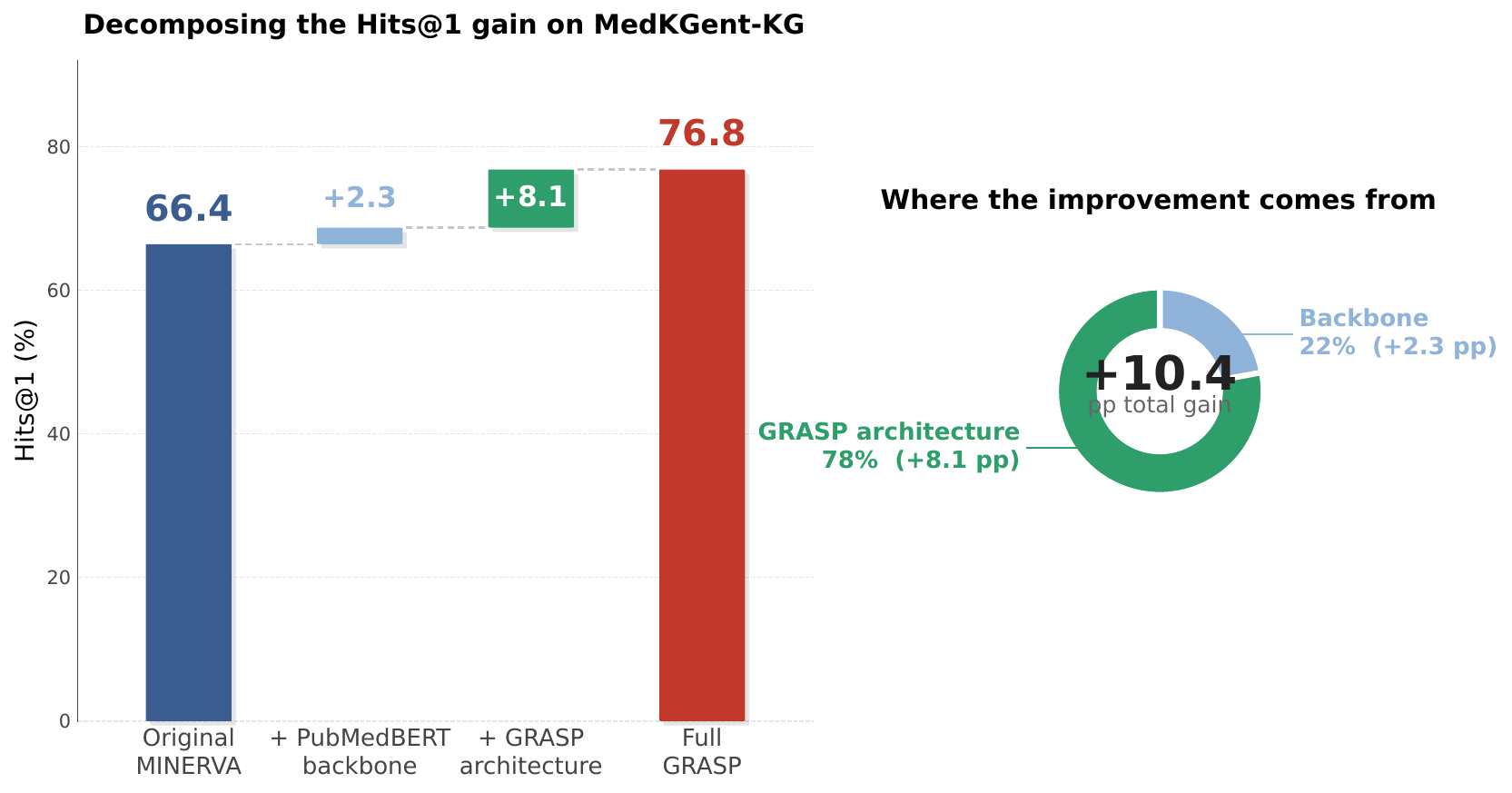}
\caption{Decomposition of the Hits@1 gain from the original MINERVA to
full GRASP on MedKGent-KG. The PubMedBERT backbone accounts for 22\%
(2.3\%) of the total 10.4\% gain, and the GRASP architecture accounts
for the remaining 78\% (8.1\%).}
\label{fig:waterfall}
\end{figure*}

Replacing the original encoder with PubMedBERT raises Hits@1 by 2.3\%
and Path Recall by 2.4\%, which leaves most of the gap to GRASP
unexplained by the backbone. Figure~\ref{fig:waterfall} decomposes the
10.4\% total Hits@1 gain over the original MINERVA. The backbone
accounts for 2.3\% (22\%), and the GRASP architecture, comprising
cross-attention scoring, the stop gate, history modulation, and the
sufficiency loop, accounts for the remaining 8.1\% (78\%). This split
holds for this dataset and backbone pair and need not transfer to other
settings.

G-Retriever and GNN-RAG differ in encoders, training objectives, and
graph representations. Their results in Table~\ref{tab:main_retrieval}
are therefore end-to-end benchmark comparisons and not causal component
comparisons.

\subsection{Ablation Studies}
\label{sec:ablation}

\begin{figure*}[!t]
\centering
\includegraphics[width=0.85\textwidth]{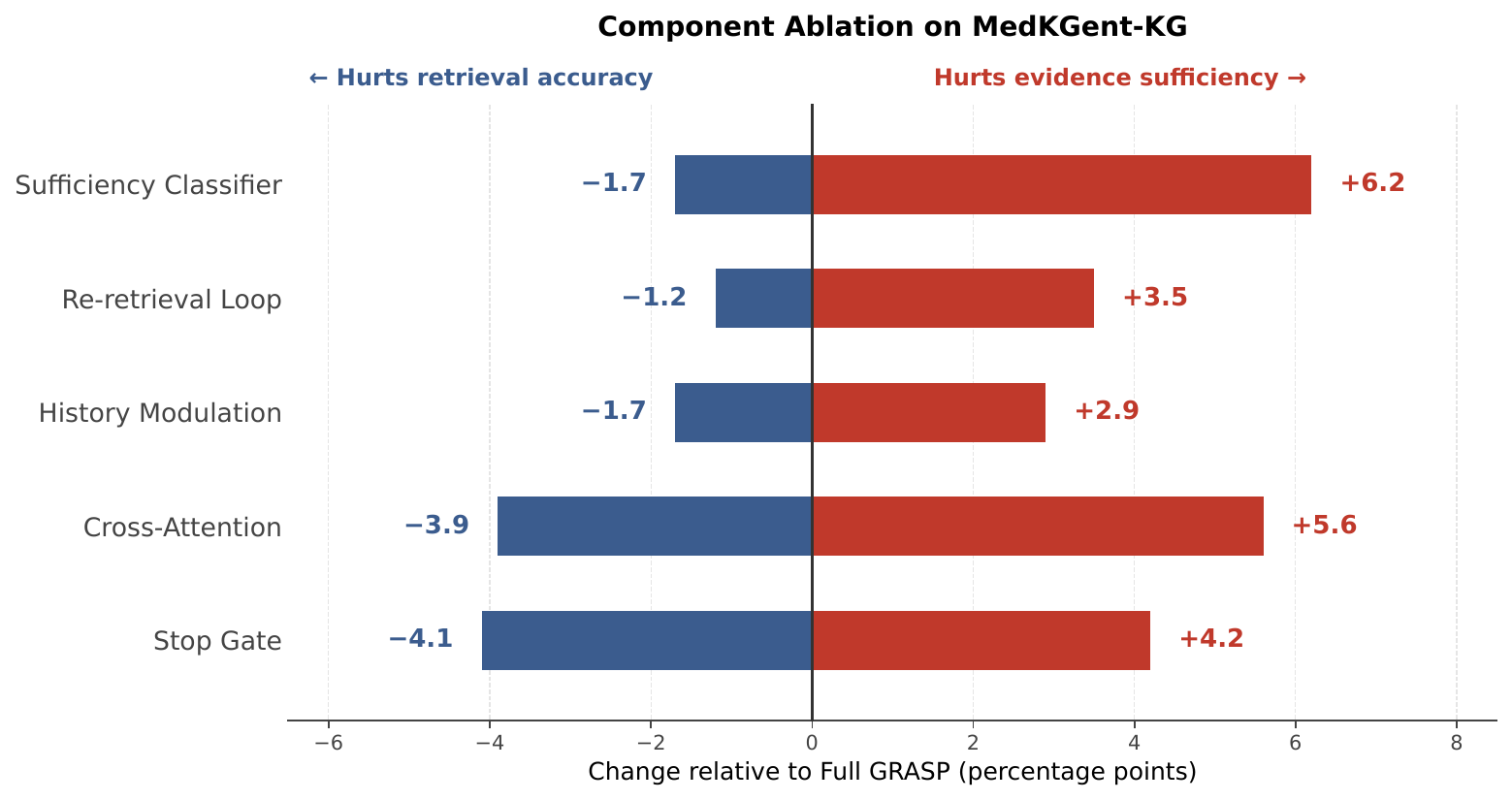}
\caption{Component ablation on MedKGent-KG, shown as the change relative
to full GRASP when each component is removed. Left bars show the change
in Hits@1 and right bars the change in Unsupported Claim Rate, both in
percent.}
\label{fig:tornado}
\end{figure*}

Each GRASP component is removed in turn and the variant is retrained
(Table~\ref{tab:ablation}). Figure~\ref{fig:tornado} shows the change in
Hits@1 and in Unsupported Claim Rate relative to full GRASP. Removing
Stage~B components mainly lowers retrieval accuracy, whereas removing the
verification components mainly raises the Unsupported Claim Rate.

\begin{table*}[!t]
\centering
\small
\setlength{\tabcolsep}{6pt}
\renewcommand{\arraystretch}{1.05}
\caption{Component ablation on MedKGent-KG. Each variant is trained
independently (mean $\pm$ standard deviation). Paired tests and
per-seed results are given in the appendix.}
\label{tab:ablation}
\begin{tabular}{lccccc}
\toprule
Variant & Hits@1 & Path Rec. & Tok. F1 & BERTScore & Unsup. $\downarrow$ \\
\midrule
w/o Cross-Attention
& $72.9\pm0.9$ & $51.8\pm1.4$ & $44.1\pm1.0$
& $0.861\pm0.004$ & $20.7\pm1.0$ \\
w/o History Modulation
& $75.1\pm0.9$ & $54.0\pm1.3$ & $46.3\pm1.0$
& $0.869\pm0.004$ & $18.0\pm0.9$ \\
w/o Stop Gate
& $72.7\pm0.9$ & $51.2\pm1.4$ & $44.0\pm1.1$
& $0.858\pm0.005$ & $19.3\pm1.0$ \\
w/o Sufficiency Classifier
& $75.1\pm0.8$ & $56.2\pm1.3$ & $48.2\pm1.0$
& $0.872\pm0.004$ & $21.3\pm1.0$ \\
w/o Re-retrieval Loop
& $75.6\pm0.9$ & $56.8\pm1.3$ & $48.6\pm1.0$
& $0.874\pm0.004$ & $18.6\pm0.9$ \\
Full GRASP
& $\mathbf{76.8\pm0.8}$ & $\mathbf{57.9\pm1.0}$ & $\mathbf{49.7\pm0.9}$
& $\mathbf{0.876\pm0.003}$ & $\mathbf{15.1\pm0.8}$ \\
\bottomrule
\end{tabular}
\end{table*}

\subsubsection{Removing Stage B Components: Cross-Attention, History Modulation, and Stop Gate}

Removing the stop gate or cross-attention causes the two largest
retrieval drops, 4.1\% and 3.9\% in Hits@1. The two variants differ by
0.2\% in Hits@1 and 0.6\% in Path Recall, within one standard
deviation, so they are not ranked. The cross-attention drop supports
token-level alignment in dense candidate neighbourhoods. Without the
stop gate, traversal depth cannot adapt to the query, which tends to
produce incomplete or over extended paths and raises the Unsupported
Claim Rate by 4.2\%. Removing history modulation has a smaller effect
that is concentrated in path quality, with Path Recall falling by 3.9\%
and Hits@1 by 1.7\%.

\subsubsection{Removing Stage C and D Components: Sufficiency Classifier and Re-retrieval Loop}

Removing the sufficiency classifier lowers Hits@1 by only 1.7\% (76.8\%
to 75.1\%) but raises the Unsupported Claim Rate by 6.2\% (15.1\% to
21.3\%), the largest increase among all components. Its Hits@1 effect is
not treated as a reliable retrieval effect, and the paired test is
reported in Appendix~\ref{sec:appendix_ablation}. 
Removing only the re-retrieval loop, with the classifier kept, lowers
Hits@1 by 1.2\% and raises the Unsupported Claim Rate by 3.5\% (15.1\%
to 18.6\%), so the loop adds a benefit beyond classification. As
Figure~\ref{fig:tornado} shows, the two verification components act
mainly on whether generation stays grounded in the retrieved path, while
cross-attention and the stop gate shift retrieval and generation to a
similar degree.

\subsection{Reliability of Sufficiency Verification}
\label{sec:sufficiency_validation}

The sufficiency classifier cannot be trained on Unsupported Claim Rate,
which is observable only after generation. It is instead trained on
structural sufficiency, defined by the correspondence between the
retrieved linear path and the annotated reasoning chain. The corruption
pool contains six operations: 75\% truncation, 50\% truncation, node
substitution, relation swap, over-extension, and reverse-path
corruption.

\begin{table}[!t]
\centering
\small
\setlength{\tabcolsep}{8pt}
\renewcommand{\arraystretch}{1.05}
\caption{Sufficiency classifier validation on the development set. F1 is
computed against the structural sufficiency label.}
\label{tab:suf_validation}
\begin{tabular}{lccc}
\toprule
Corruption & Precision & Recall & F1 \\
\midrule
Truncation 75\% & 0.89 & 0.91 & 0.90 \\
Truncation 50\% & 0.92 & 0.93 & 0.92 \\
Node substitution & 0.85 & 0.84 & 0.85 \\
Relation swap & 0.83 & 0.82 & 0.83 \\
Over-extension & 0.82 & 0.81 & 0.81 \\
Reverse-path & 0.90 & 0.88 & 0.89 \\
\midrule
Overall synthetic pool & 0.87 & 0.87 & 0.87 \\
Hard negatives & 0.79 & 0.77 & 0.78 \\
\bottomrule
\end{tabular}
\end{table}

The classifier reaches an F1 of 0.87 on the synthetic pool but 0.78 on
hard negatives generated by Stage~B. Hard negatives contain errors from
the learned traversal policy and not from predefined corruption rules,
so the synthetic score overstates deployment performance. The
classifier also relies on Stage~B confidence statistics such as action
margins and entropy, which are informative when the scorer is correct
but can be over-confident on wrong actions, so it is not an independent
oracle. Each corruption type has few development examples, and the
per-type values carry no confidence intervals.

The structural proxy is useful only if it tracks the support of
generated claims. We correlate the proxy score with per-question
Path-Supported Atomic Fact Rate (PSAFR), which measures the
proportion of generated atomic facts supported by the retrieved
evidence path. PSAFR is computed from the same held-out generations,
claim decomposition, and NLI decisions as
Table~\ref{tab:main_generation}, so that
\begin{equation*}
\mathrm{Unsupported}
= 1-\mathrm{PSAFR}
= \mathrm{PNR}+\mathrm{PCR},
\end{equation*}
where PNR denotes Path-Neutral Rate, measuring the proportion of
claims neither supported nor contradicted by the retrieved path,
and PCR denotes Path-Contradiction Rate, measuring the proportion
of claims contradicted by the retrieved path.
The correlation is positive on all three datasets
(Table~\ref{tab:proxy_validation}).
The lower value on PMC agrees with its automatically extracted
graph and the absence of path annotations.
The proxy is therefore a structural predictor of generation support
and not a substitute for generation-level evaluation.
The analysis covers the full validation distribution and may not
hold on failure subsets such as hard negatives, whose upstream
errors the corruption pool may not represent.
\begin{table*}[!t]
\centering
\small
\setlength{\tabcolsep}{6pt}
\renewcommand{\arraystretch}{1.05}
\caption{Validation of the structural sufficiency proxy against
generation-level PSAFR. Generated PSAFR is computed from the held-out
generations used in Table~\ref{tab:main_generation}, and $r$ is
Pearson's correlation between the structural proxy score and
per-question PSAFR.}
\label{tab:proxy_validation}
\begin{tabular}{lcccc}
\toprule
Dataset & Generated PSAFR & Proxy $r$ & 95\% CI & $p$ \\
\midrule
RiTeK
& $88.4\%$ & 0.83 & [0.76, 0.88] & $<0.001$ \\
MedKGent-KG
& $84.9\%$ & 0.81 & [0.72, 0.87] & $<0.001$ \\
PMC
& $84.6\%$ & 0.73 & [0.62, 0.82] & $<0.001$ \\
\bottomrule
\end{tabular}
\end{table*}

\subsection{Re-retrieval Loop: Which Failures Can Be Recovered}
\label{sec:reroute_analysis}

The loop preserves the traversal prefix before the selected hop and
replaces the action at the lowest-margin location, so it performs local
correction and not global backtracking. On datasets with gold paths,
failed paths are assigned post hoc to three categories, using the
observed traversal prefix and the gold path.

\begin{itemize}
    \item \emph{Edge-selection error}: the traversal follows the gold
    prefix up to the selected hop but chooses a different action, which
    tends to be locally reroutable.
    \item \emph{Premature stopping}: the traversed prefix is correct but
    the stop decision comes before the gold path is complete, so
    replacing one action cannot force another hop.
    \item \emph{Wrong region entry}: the traversal leaves the gold path
    before the weakest margin hop, so replacing a later action cannot
    restore the lost prefix.
\end{itemize}

\begin{table*}[!t]
\centering
\small
\setlength{\tabcolsep}{6pt}
\renewcommand{\arraystretch}{1.05}
\caption{Post-hoc analysis of re-retrieval failures on datasets with gold
paths. Frequencies are computed before rounding, and recovery requires an
exact match with the annotated gold chain after re-retrieval. The overall
rate is the frequency-weighted average of the category-level rates.}
\label{tab:reroute}
\begin{tabular}{lccc}
\toprule
Failure type & Frequency & Recovery & Interpretation \\
\midrule
Edge-selection error
& 46.9\% & 64.8\% & Locally recoverable \\
Premature stopping
& 29.1\% & 17.6\% & Requires extension \\
Wrong-region entry
& 24.0\% & 12.4\% & Requires earlier correction \\
\midrule
Overall
& 100\% & 38.5\% & Partial recovery \\
\bottomrule
\end{tabular}
\end{table*}

The overall recovery rate of 38.5\% is far below the 64.8\% for
edge-selection errors. This fits the design of the loop: it repairs a
locally wrong action but cannot extend a prematurely stopped path or
restore a prefix lost at entry. Recovery requires an exact match with the
gold chain, which is stricter than the acceptance decision of Stage~C
reported in Appendix~C, since a path can be accepted as structurally
sufficient without matching the gold chain. The two rates therefore
measure different quantities, and their gap reflects the imperfection of
the classifier discussed above. The loop lowers the Unsupported Claim
Rate through failures that can be corrected locally and through the
rejection of paths that remain insufficient, not through recovering every
failure.

\subsection{Re-retrieval Loop: Sensitivity to $I_{\max}$ and $\tau_{\mathrm{cls}}$}
\label{sec:loop_sensitivity}

We vary the maximum number of re-retrieval attempts
$I_{\max}\in\{1,2,3,4\}$ and the classifier threshold
$\tau_{\mathrm{cls}}\in\{0.40,0.45,0.50,0.55\}$.

\begin{table*}[!t]
\centering
\small
\setlength{\tabcolsep}{8pt}
\renewcommand{\arraystretch}{1.05}
\caption{Sensitivity of the sufficiency loop on MedKGent-KG. Each cell
reports Unsupported Claim Rate followed by Hits@1 in parentheses. The
default configuration, $I_{\max}=3$ and $\tau_{\mathrm{cls}}=0.50$,
matches the full GRASP row in Tables~\ref{tab:main_retrieval}
and~\ref{tab:main_generation}.}
\label{tab:loop_sensitivity}
\begin{tabular}{lcccc}
\toprule
$I_{\max}\backslash\tau_{\mathrm{cls}}$
& 0.40 & 0.45 & 0.50 & 0.55 \\
\midrule
1
& 19.0 (73.9) & 18.1 (74.6) & 17.3 (75.3) & 16.8 (75.0) \\
2
& 17.2 (75.8) & 16.4 (76.3) & 15.7 (76.6) & 15.3 (76.2) \\
3
& 16.6 (76.5) & 15.6 (76.8) & \textbf{15.1 (76.8)}
& 14.9 (76.3) \\
4
& 16.4 (76.5) & 15.3 (76.8) & 15.0 (76.8)
& 14.8 (76.0) \\
\bottomrule
\end{tabular}
\end{table*}

\begin{figure*}[!t]
\centering
\includegraphics[width=0.85\textwidth]{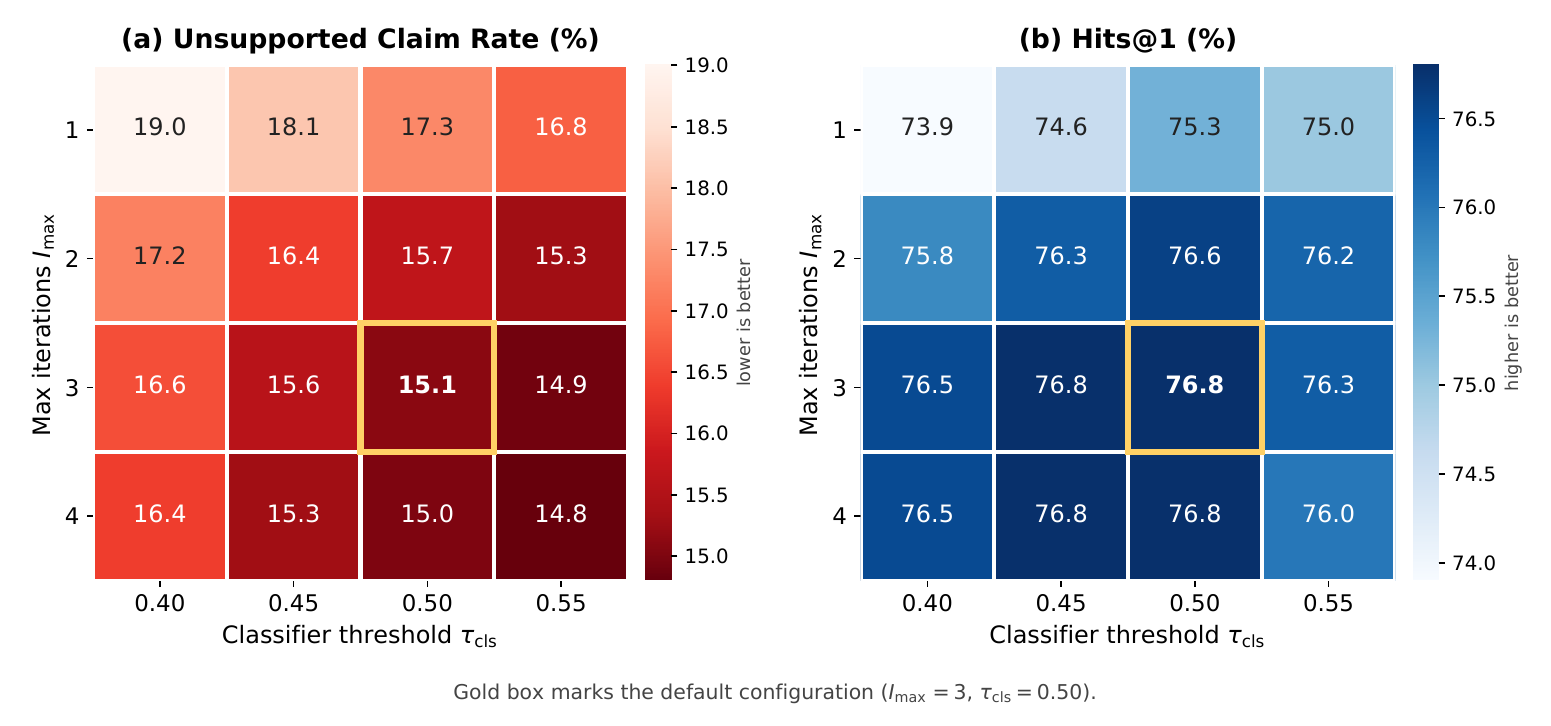}
\caption{Sensitivity of GRASP to the maximum number of re-retrieval
attempts $I_{\max}$ and the classifier threshold
$\tau_{\mathrm{cls}}$ on MedKGent-KG. Subfigure (a) shows Unsupported
Claim Rate (lower is better) and subfigure (b) shows Hits@1 (higher is
better). The gold-bordered cell marks the default configuration.}
\label{fig:heatmap}
\end{figure*}

At the default threshold, raising $I_{\max}$ from one to three lowers the
Unsupported Claim Rate from 17.3\% to 15.1\% and raises Hits@1 from
75.3\% to 76.8\%. A fourth iteration changes the Unsupported Claim Rate by
only 0.1\% and leaves Hits@1 unchanged, so it gives no evidence for a
larger budget. A higher $\tau_{\mathrm{cls}}$ makes the classifier more
conservative and lowers the Unsupported Claim Rate, but Hits@1 falls
beyond 0.50 (76.8\% to 76.3\% at $I_{\max}=3$) as borderline paths are
rejected. The default setting reaches the highest Hits@1 and stays within
0.3\% of the lowest Unsupported Claim Rate in the grid
(Figure~\ref{fig:heatmap}).

\subsection{Candidate Density: Where Token-Level Scoring Helps}
\label{sec:density_analysis}

Token-level cross-attention is designed for neighbourhoods in which many
plausible edges compete. To separate this factor from cross-dataset
differences, queries in MedKGent-KG are grouped by the out-degree of
their starting entity, and GRASP is compared with \mpb within each
bin.

\begin{table}[!t]
\centering
\small
\setlength{\tabcolsep}{8pt}
\renewcommand{\arraystretch}{1.05}
\caption{Candidate-density analysis on MedKGent-KG (Hits@1). The gap is
GRASP minus \mpb.}
\label{tab:density}
\begin{tabular}{lccc}
\toprule
Out-degree & MINERVA-PBERT & GRASP & Gap \\
\midrule
$[5,10)$
& 70.8 & 75.1 & $+4.3\%$ \\
$[10,20)$
& 68.9 & 76.1 & $+7.2\%$ \\
$[20,50)$
& 65.4 & 76.8 & $+11.4\%$ \\
$[50,\infty)$
& 61.8 & 77.4 & $+15.6\%$ \\
\bottomrule
\end{tabular}
\end{table}

\begin{figure*}[!t]
\centering
\includegraphics[width=0.85\textwidth]{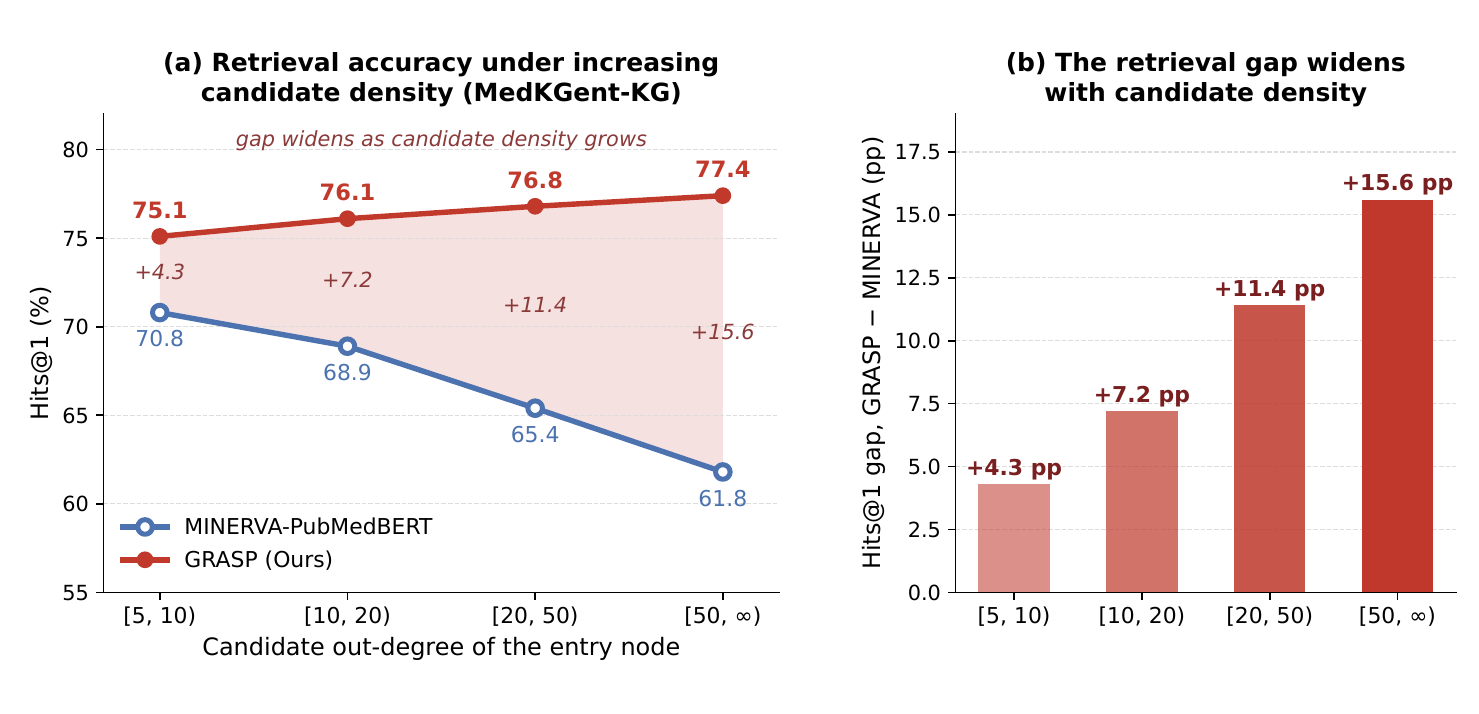}
\caption{Retrieval accuracy as a function of candidate out-degree on
MedKGent-KG. Subfigure (a) shows Hits@1 of \mpb and GRASP across four
out-degree bins, and subfigure (b) shows the Hits@1 gap between them,
which widens with candidate density.}
\label{fig:density}
\end{figure*}

As out-degree grows, the Hits@1 of \mpb falls from 70.8\% to 61.8\%,
while that of GRASP rises slightly from 75.1\% to 77.4\%
(Figure~\ref{fig:density}). The gap widens from 4.3\% in the sparsest bin
to 15.6\% in the densest, in line with the motivation for token-level
cross-attention: when more locally plausible edges compete for the same
node, a pooled query representation is less discriminative. The test set
has 100 queries, so each bin holds only a fraction of them and no
confidence intervals are reported. The trend is therefore an association
within one dataset and not a causal estimate.

\subsection{Efficiency}
\label{sec:efficiency}

Stage~C and the re-retrieval loop add computation.
Table~\ref{tab:efficiency} reports the average inference latency and the
number of re-retrieval attempts.

\begin{table}[!t]
\centering
\small
\setlength{\tabcolsep}{2pt}
\renewcommand{\arraystretch}{1.05}
\caption{Inference efficiency on MedKGent-KG. Latency is measured per
query and includes retrieval, sufficiency checking, and generation
setup.}
\label{tab:efficiency}
\begin{tabular}{@{}>{\raggedright\arraybackslash}p{1.25cm}cccc@{}}
\toprule
Model & Latency (ms) & Retries & LLM tokens & GPU memory \\
\midrule
\mpb
& 96 & 0 & 1.00$\times$ & 6.8 GB \\
GRASP w/o Loop
& 141 & 0 & 1.02$\times$ & 7.1 GB \\
GRASP
& 226 & 0.31 & 1.07$\times$ & 7.4 GB \\
\bottomrule
\end{tabular}
\end{table}

Most of the added latency comes from the cross-attention scorer and the
sufficiency evaluation. The loop adds 85 ms per query on average and
lowers the Unsupported Claim Rate by 3.5\% (Table~\ref{tab:ablation}),
and it is invoked only when a path is rejected.

\subsection{Summary of the Experimental Evidence}
\label{sec:experimental_discussion}

The experiments support three conclusions, each tied to a design choice
of GRASP. First, token-level scoring with adaptive stopping improves path
retrieval. GRASP raises Path Recall by 13.9\% and 13.0\% over MINERVA,
the controlled comparison attributes 78\% of the Hits@1 gain to the
architecture and not to the backbone, and the gap to \mpb widens from
4.3\% to 15.6\% as candidate density grows.

Second, adaptive stopping and sufficiency verification play different
roles. The stop gate sets how long the traversal continues, whereas the
sufficiency classifier judges the completed path. Removing the stop gate
mainly lowers retrieval accuracy, and removing the classifier raises the
Unsupported Claim Rate by 6.2\%, so the two components are not
interchangeable.

Third, the re-retrieval loop reduces unsupported claims but repairs only
local failures. Recovery is 64.8\% for edge-selection errors and 38.5\%
overall, so the loop is targeted local correction and not general
backtracking. The structural proxy correlates with generation-level
PSAFR ($r=0.73$ to $0.83$), which supports its use as a training signal,
but its weaker performance on hard negatives and on PMC shows that it
does not replace generation-level verification.

\section{Conclusion}
\label{sec:conclusion}

This work introduced GRASP, a graph-based retrieval framework that
combines query-adaptive edge scoring, learned traversal termination,
and path-sufficiency verification. Experiments across RiTeK, MedKGent-KG,
and PMC show improvements in retrieval quality and reductions in
Unsupported Claim Rate over the evaluated learned baselines. The
controlled MINERVA comparison supports the contribution of the proposed
retrieval mechanisms beyond the choice of encoder, while the component
analysis indicates that edge scoring, stopping, and sufficiency
verification contribute in complementary ways. The blind-tail experiment
further shows that the retrieval advantage persists when the target
entity is removed from the query. These findings support the value of
integrating adaptive traversal with path-level verification, although
they do not establish that the approach can resolve all retrieval
failures or that structural path sufficiency guarantees factual
correctness.

\subsection{Limitations}
\label{sec:limitations}

GRASP has three main limitations. First, the structural proxy used to
train the sufficiency classifier only approximates generation-level
evidence support, and its reliance on upstream retrieval confidence
makes it vulnerable to confidently incorrect actions. Second, the
re-retrieval mechanism modifies a low margin action locally and may
fail to recover from premature stopping or entry into an incorrect
graph region. Third, PMC lacks gold path annotations and uses an
automatically constructed graph that may contain missing or spurious
relations; moreover, automatic NLI evaluation only approximates human
judgement. Consequently, the classifier's reliability, the recovery
capability of re-retrieval, and the interpretation of Unsupported
Claim Rate remain limited.

\subsection{Future Work}
\label{sec:future_work}

Future work will address these limitations by evaluating the structural
sufficiency proxy on larger, naturally occurring sets of retrieval
failures and calibrating the Stage~B confidence features under
distribution shift. The re-retrieval mechanism could be extended with
explicit backtracking, path extension, and abstention actions to better
handle premature stopping and incorrect-region entry. Further work
could also establish more controlled comparisons with G-Retriever and
GNN-RAG under a shared encoder and training protocol, and validate
automatic NLI based support assessments through human annotation on a
larger, stratified biomedical sample.

\bibliographystyle{elsarticle-num}
\bibliography{cas-refs}
\appendix
\renewcommand{\thesection}{Appendix \Alph{section}}

\section{Additional Statistical Results}
\label{sec:appendix_statistics}

\begin{table*}[htbp]
\centering
\small
\caption{Per-seed retrieval results for the main models. Hits@1 is
reported as a whole-number percentage because the held-out test set
contains exactly 100 instances per benchmark
(Section~\ref{sec:experimental_protocol}), so each seed's Hits@1 is
necessarily a multiple of 1\%. MRR and Path Recall are continuous
averages over the 100 queries and are not subject to this
constraint.}
\label{tab:per_seed_retrieval}
\begin{tabular}{lccccc}
\toprule
Dataset & Model & Seed & Hits@1 & MRR & Path Recall \\
\midrule
RiTeK & MINERVA & 1 & 74 & 0.778 & 52.3 \\
RiTeK & MINERVA & 2 & 75 & 0.783 & 53.5 \\
RiTeK & MINERVA & 3 & 75 & 0.786 & 54.2 \\
RiTeK & MINERVA & 4 & 76 & 0.789 & 54.7 \\
RiTeK & MINERVA & 5 & 76 & 0.789 & 54.3 \\
\midrule
RiTeK & GRASP & 1 & 84 & 0.874 & 66.4 \\
RiTeK & GRASP & 2 & 85 & 0.878 & 68.2 \\
RiTeK & GRASP & 3 & 85 & 0.881 & 67.3 \\
RiTeK & GRASP & 4 & 86 & 0.883 & 68.9 \\
RiTeK & GRASP & 5 & 86 & 0.884 & 67.9 \\
\midrule
MedKGent-KG & MINERVA & 1 & 65 & 0.701 & 43.3 \\
MedKGent-KG & MINERVA & 2 & 66 & 0.708 & 46.3 \\
MedKGent-KG & MINERVA & 3 & 66 & 0.712 & 44.1 \\
MedKGent-KG & MINERVA & 4 & 67 & 0.715 & 46.0 \\
MedKGent-KG & MINERVA & 5 & 68 & 0.719 & 44.8 \\
\midrule
MedKGent-KG & GRASP & 1 & 76 & 0.799 & 56.5 \\
MedKGent-KG & GRASP & 2 & 77 & 0.804 & 59.0 \\
MedKGent-KG & GRASP & 3 & 77 & 0.808 & 58.1 \\
MedKGent-KG & GRASP & 4 & 78 & 0.810 & 58.7 \\
MedKGent-KG & GRASP & 5 & 76 & 0.814 & 57.2 \\
\bottomrule
\end{tabular}
\end{table*}

The MedKGent-KG baseline reported here is the original-backbone
MINERVA used in Table~\ref{tab:main_retrieval}, not the
PubMedBERT-backbone variant used only in the controlled comparison of
Table~\ref{tab:controlled}. The two should not be conflated: they
differ only in backbone, but are trained and evaluated as separate
configurations.

With five integer valued seeds, the sample mean of Hits@1 is
necessarily a multiple of $0.2$\%. For MedKGent-KG
GRASP the five seed mean is $76.8$, which is the value reported in
Table~\ref{tab:main_retrieval}. For the four model dataset
combinations listed above, the means and sample standard deviations
in Table~\ref{tab:main_retrieval} are computed directly from these
per-seed values.

\section{Paired Significance Tests and Additional Ablation Statistics}
\label{sec:appendix_ablation}

Section~\ref{sec:ablation} reports mean and standard deviation for
each ablation variant but defers paired significance testing to this
appendix. Table~\ref{tab:appendix_ablation_sig} reports, for each
variant, the paired bootstrap difference from Full GRASP on Hits@1
and Unsupported Claim Rate, using the same question-level resampling
procedure as Table~\ref{tab:significance}.

\begin{table*}[htbp]
\centering
\small
\caption{Paired bootstrap tests for the component ablation on
MedKGent-KG. $\Delta$ is the variant's value minus the Full GRASP
value; a positive $\Delta$ on Unsupported Claim Rate indicates a
worse (higher) rate.}
\label{tab:appendix_ablation_sig}
\begin{tabular}{lcccc}
\toprule
Variant & $\Delta$ Hits@1 (pp) & $p$ & $\Delta$ Unsup. (pp) & $p$ \\
\midrule
w/o CA     & -3.9 & $<0.001$ & +5.6 & $<0.001$ \\
w/o HM     & -1.7 & $0.006$  & +2.9 & $0.003$ \\
w/o Stop   & -4.1 & $<0.001$ & +4.2 & $<0.001$ \\
w/o Suf    & -1.7 & $0.014$  & +6.2 & $<0.001$ \\
w/o Loop   & -1.2 & $0.038$  & +3.5 & $<0.001$ \\
\bottomrule
\end{tabular}
\end{table*}

The Hits@1 difference for the sufficiency classifier (w/o Suf) is
only marginally significant at the conventional 0.05 threshold. This
supports the caution expressed in Section~\ref{sec:ablation}: the
1.7-point Hits@1 change should not be treated as a reliable retrieval
effect on its own, while the corresponding change in Unsupported
Claim Rate is significant well beyond that threshold. The remaining
variants show significant differences on both metrics.

For completeness, Table~\ref{tab:appendix_ablation} in Appendix X reproduces the
mean $\pm$ standard deviation values from Table~\ref{tab:ablation}.

\begin{table*}[htbp]
\centering
\small
\caption{Component ablation on MedKGent-KG, mean $\pm$ standard
deviation over five independent training runs. Values match
Table~\ref{tab:ablation} and are repeated here for reference alongside
the significance tests above.}
\label{tab:appendix_ablation}
\begin{tabular}{lccccc}
\toprule
Variant & Hits@1 & Path Rec. & Tok. F1 & BERTScore & Unsup. \\
\midrule
w/o CA
& $72.9\pm0.9$ & $51.8\pm1.4$ & $44.1\pm1.0$
& $0.861\pm0.004$ & $20.7\pm1.0$ \\
w/o HM
& $75.1\pm0.9$ & $54.0\pm1.3$ & $46.3\pm1.0$
& $0.869\pm0.004$ & $18.0\pm0.9$ \\
w/o Stop
& $72.7\pm0.9$ & $51.2\pm1.4$ & $44.0\pm1.1$
& $0.858\pm0.005$ & $19.3\pm1.0$ \\
w/o Suf
& $75.1\pm0.8$ & $56.2\pm1.3$ & $48.2\pm1.0$
& $0.872\pm0.004$ & $21.3\pm1.0$ \\
w/o Loop
& $75.6\pm0.9$ & $56.8\pm1.3$ & $48.6\pm1.0$
& $0.874\pm0.004$ & $18.6\pm0.9$ \\
Full
& $\mathbf{76.8\pm0.8}$ & $\mathbf{57.9\pm1.0}$
& $\mathbf{49.7\pm0.9}$ & $\mathbf{0.876\pm0.003}$
& $\mathbf{15.1\pm0.8}$ \\
\bottomrule
\end{tabular}
\end{table*}

\section{Sufficiency Loop Behaviour}
\label{sec:appendix_loop}

\begin{table*}[htbp]
\centering
\small
\caption{Empirical behaviour of the re-retrieval loop. Trigger Rate
is the fraction of test queries for which Stage C rejects the
initial path at least once. Retry~1 Accepted and Retry~2 Accepted are
the fraction of triggered cases accepted by Stage C after the first
and second re-retrieval attempt, respectively; Fallback is the
remaining fraction that exhausts the iteration budget $I_{\max}=3$
and falls back to the highest-margin path (Section~4.3). The three
right-hand columns sum to 100\% of the triggered cases in each row.}
\label{tab:appendix_loop}
\begin{tabular}{lcccc}
\toprule
Dataset & Trigger Rate & Retry 1 Accepted
& Retry 2 Accepted & Fallback \\
\midrule
RiTeK
& 18.4\% & 64.1\% & 31.2\% & 4.7\% \\
MedKGent-KG
& 22.7\% & 59.8\% & 33.7\% & 6.5\% \\
PMC
& 24.8\% & 55.6\% & 35.1\% & 9.3\% \\
\bottomrule
\end{tabular}
\end{table*}

``Accepted'' here means that Stage C's sufficiency classifier judges
the revised path adequate, which is the operational stopping
condition used at inference time (Section~4.3). This is a weaker
criterion than the recovery rate reported in
Table~\ref{tab:reroute}, which additionally requires the revised path
to match the annotated gold chain exactly. A path can therefore be
accepted here without being recovered in the sense of
Table~\ref{tab:reroute} the gap between the two statistics reflects
the classifier's imperfect correlation with true sufficiency,
discussed in Section~\ref{sec:sufficiency_validation}, rather than an
inconsistency between the two tables. The higher fallback rate on PMC
is consistent with its noisier, automatically extracted graph, which
makes an exact structural match harder to reach within the fixed
iteration budget.

\section{Proxy and NLI Verification}
\label{sec:appendix_nli}

\begin{table*}[htbp]
\centering
\small
\caption{Generation-level verification and human validation. The
human validation sample is stratified across datasets.}
\label{tab:appendix_nli}
\begin{tabular}{lcccc}
\toprule
Dataset & Claims & Entail P/R/F1 & Neutral P/R/F1
& Contradict P/R/F1 \\
\midrule
RiTeK
& 70 & 0.88/0.85/0.86 & 0.80/0.83/0.81 & 0.78/0.75/0.76 \\
MedKGent-KG
& 70 & 0.86/0.84/0.85 & 0.79/0.81/0.80 & 0.76/0.73/0.74 \\
PMC
& 80 & 0.83/0.79/0.81 & 0.75/0.78/0.76 & 0.71/0.69/0.70 \\
\bottomrule
\end{tabular}
\end{table*}

The human validation yields an overall Cohen's $\kappa$ of 0.78. These
results are used only to assess the automatic verifier, and are not
used to redefine the main Unsupported Claim Rate. They provide the
supporting evidence for treating Unsupported Claim Rate as an
approximation to, rather than an exact measurement of, clinical
hallucination, a caveat raised in the paper's limitations discussion.
The lower per class scores on PMC are consistent with its longer,
less structured reference answers, which make claim decomposition and
NLI judgment comparatively harder.

\end{document}